\documentclass[aps, prl, reprint, showpacs, twocolumn, 10pt, groupedaddress]{revtex4-2}

\usepackage{graphicx, amsmath, dsfont, dcolumn, bm, times, color, braket, amssymb}
\usepackage{subcaption}

\begin{document}

\title{Abundance of Weyl points in semiclassical multi-terminal superconducting nanostructures}

\author{Hristo Barakov}
\affiliation{Kavli Institute of Nanoscience, Delft University of Technology, 2628 CJ Delft, The Netherlands}
\author{Yuli V. Nazarov}
\affiliation{Kavli Institute of Nanoscience, Delft University of Technology, 2628 CJ Delft, The Netherlands}

\date{\today}

\begin{abstract}
We show that the quasi-continuous gapless spectrum of Andreev bound states in multi-terminal semi-classical superconducting nanostructures exhibits a big number of topological singularities. We concentrate on Weyl points in a 4-terminal nanostructure, compute their density and correlations in 3D parameter space for a universal RMT model as well as for the concrete nanostructures described by the quantum circuit theory. We mention the opportunities for experimental observation of the effect in a quasi-continuous spectrum.  
\end{abstract}

\maketitle

\let\oldvec\vec
\renewcommand{\vec}[1]{\ensuremath{\boldsymbol{#1}}}

The topological properties of quantum spectra in condensed matter systems got considerable attention in the past decade and are still under active consideration \cite{gen1,gen2,gen3,gen4}. A large research field that has been formed thereby addresses gapped phases of insulators \cite{Topinsulators} and superconductors \cite{TopSuperconductors} characterized by globally defined  topological numbers, and the edge modes \cite{bernevig2013topological} at the interfaces separating such phases. In addition to this, the spectra can exhibit topological singularities in the form of level crossings where the topological charge is defined at the singularity rather than globally. The simplest example of such singularity is a Weyl point (WP) \cite{Weyl} corresponding to crossing of two levels in a point in 3D space of parameters. Physical realizations of WPs include special points in the bandstructure of 3D solids \cite{WeylSemimetals}, spectra of polyatomic molecules \cite{faure} and nanomagnets \cite{werns2}, quantum transport systems \cite{leone2008cooper}. 

The occurrence of WPs have been recently predicted in the spectrum of Andreev bound states (ABS) of generic 4-terminal superconducting nanostructures \cite{grenoble} where the 3D parameter space is formed by three independent superconducting phases of the terminals. Most important WPs are the crossings at {\it zero energy} that define the topology of the ground state. These WPs in 3D give rise to 2D global Chern numbers that are directly manifested as quantized transconductances of the nanostructure. The ideal periodicity of the space of superconducting phases allows to model higher-dimensional bandstructures with the multi-terminal superconducting nanostructures (MTSN). These ideas resulted in outburst of theoretical \cite{thy1,thy2,thy3,thy4,thy5,thy6,thy7} and experimental \cite{exp1,exp2,exp3,exp4,exp5}  activities in the field of MTSN. 

A separate recent development concerned semiclassical MTSN where a big number of ABS form a quasi-continuous spectrum. It has been predicted \cite{Padurariu} that this spectrum can be either gapped or gapless depening on speficics of the MTSN and the point in the space of the superconducting phases. A specific topology can be introduced in semiclassical MTSN. It has been discovered and confirmed experimentally \cite{omega1,omega2} that the gapped phases are characterised by topological numbers, and the gapless phase is explained by topological protection of these numbers \cite{Yokoyama2017}. The protection-unprotection transition has been discussed in this context \cite{Xiaoli}.  
\begin{figure}[tb]
\begin{center}
\includegraphics[width=0.92\linewidth]{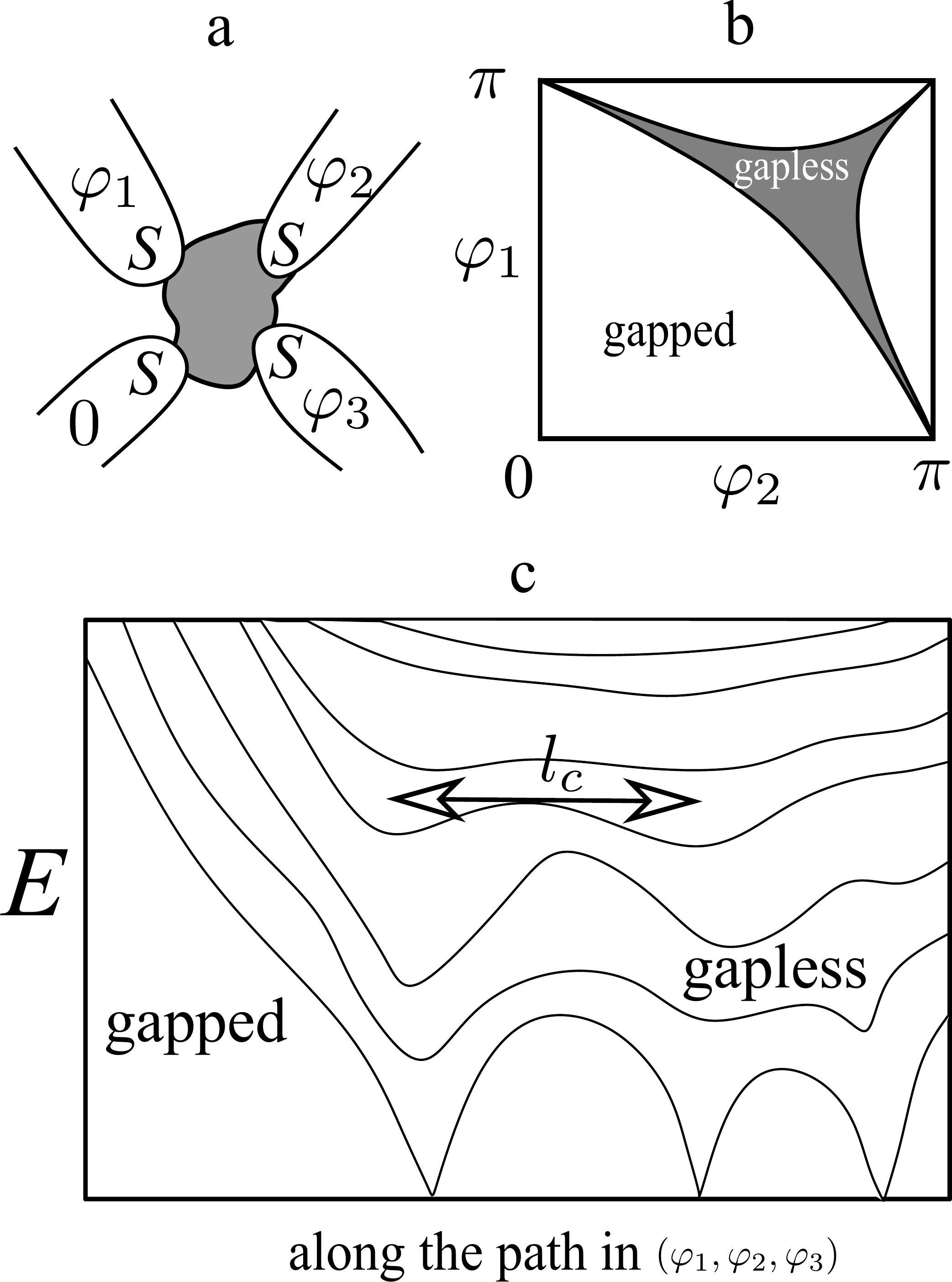}
\end{center}
\caption{Weyl points in semi-classical MTSN. a. 4-terminal semiconducting nanostructure, three independent phases forming a parameter space. b. The domains of gapped and gapless phases at $\varphi_3=0$. c. The discrete spectrum near the boundary of gapped and gapless domains plotted along a path in 3D parametric space that goes via the WPs. The distance between the WP's is of the order of the local value of $l_c$, a parameter governing the universal parametric correlations in the corresponding random matrix ensemble.  }
\label{fig:fig1}
\end{figure}

In this Letter, we analyse the gapless spectrum at the level of discrete states and reveal the abundance of zero-energy topological singularities (Fig. \ref{fig:fig1} c). In 4-terminal structures, those are isolated WPs separated by a typical distance $l_c \simeq (G/G_Q)^{-1/2} \ll 2\pi$. ($G$ is a typical conductance of the nanostructure, $G_Q \equiv e^2/\pi\hbar$). The positions of WPs are random determined by details of electron interference in the structure, while their averaged density and its correlations are determined by the structure design. We relate the density of WPs to the parameter $l_c$ governing the universal parametric correlations \cite{universal1,universal2} in random matrix ensembles, show how to compute this density for concrete nanostructures, investigate the density correlations manifested as the transconductace of the structure, and shortly discuss the opportunities of experimental detection of the WPs in semiclassical MTSN's.

Let us start with qualitative estimations. Given a 4-terminal nanostructure of a typical conductance $G$ one expects $\simeq G/G_Q$, $G_Q \equiv e^2/\pi\hbar$, conduction channels, and, correspondingly, $\simeq G/G_Q$ discrete Andreev bound states affected by superconductivity. This estimation is valid both for "short" nanostructures with the typical size smaller than the superconducting coherence length, where these levels are spread in energy interval $\Delta$, $\Delta$ being the superconducting coherence length, and "large" nanostructure where these levels are concentrated in a much smaller energy interval $E_{\rm{Th}} \simeq (G/G_Q)\delta_S$, $\delta_S$ being the mean level spacing in the normal state. The energies of these levels depend on the 3 superconducting phases. Owing to periodicity in phases, the spectrum is to be considered in a Brillouin zone (BZ) of the size $\simeq 2\pi$.  The RMT of parametric correlations suggests that the level energies wiggle randomly. The energies change at the scale of the level spacing at  a typical distance $l_c$ in the parameter space \cite{universal1,universal2}. This distance is determined from comparison of the mean fluctuations of the derivatives of the energies with respect to the parameters and this level spacing. For our situation, the estimation $l_c \simeq \sqrt{G/G_Q}$ in the space of phases holds for both long and short nanostructures. To understand WP's we concentrate on the level that is closest to zero energy. Upon wiggling, it will reach zero at a typical distance of the order of $l_c$. Therefore, the total number of WP's in the Brillouin zone can be estimated as $N_w \simeq (l_c)^{-3} \simeq (G/G_Q)^{3/2}$. 

Our detailed results (See Fig. \ref{fig:fig3}) indeed give \begin{equation}
\label{eq:density}
N_w = A (G/G_Q)^{3/2}
\end{equation} for the cross-like structures with the arm conductances $G$ where $A=0.40$ for the ballistic conductor and $A=0.16$ for the diffusive one. The dimensionless coefficient $A < 1$, this is explained by a rather small fraction of BZ volume taken by the gapless phase ($25 \%$ for ballistic and $18 \%$ for diffusive cross). 

As it was shown in \cite{grenoble} the transconductance of the structure is defined by a Chern number of a plane traversing the BZ. The difference of two Chern numbers corresponding to two different planes is given by the total charge of the WP's enclosed between the planes. A naive estimation of the variance of this difference would be the number of WP's enclosed, $ \langle\langle(C_1-C_2)  \rangle \rangle\simeq N_w \simeq d l_c^{-3}$, $2\pi \ll d \gg \l_c$ being the separation of the planes. This estimation would hold for randomly placed uncorrelated WP's. However, they do correlate similar to ions in an electroneutral gas: a charge of a WP is screened by other points at the distance of the order of their separation, that is, of $l_c$. Therefore, only WP's at a distance $\simeq l_c$ contribute to the fluctuation of the Chern number and $ \langle\langle(C_1-C_2) \rangle \rangle \simeq l_c^{-2} \simeq (G/G_Q)$. A typical transconductance is thus $\simeq \sqrt{G_Q G}$. 

Our quantitative results are obtained in the course of three activities: A. we study numerically a generic RM model to relate the density of WPs to $l_c^{-3}$ and quantify the correlations of WP's. B. we develop a theory to compute $l_c^{-3}(\vec{\phi})$ for any MTSN described by the quantum circuit theory \cite{QuantumTransport} and derive concrete expressions for a single-node circuit. C. We find numerically the positions of WP's in the ballistic cross junction (Fig. 3 a) to prove the consistency of the results obtained in the activities A and B. The details of all activities are given in \cite{supplmat}.

{\it Activity A.} The studies of statistics of spectral crossings have been pioneered by Wilkinson et al. \cite{Wilkinson1,Wilkinson2,Wilkinson3}. They introduced a convenient RMT model in a 3D parameter space $\{\phi_i\}$,
\begin{equation}
H(\vec{\phi})=\sum_{i=1}^{3} \left(\sin \phi_i X_i + \cos\phi_i Y_i\right)
\end{equation}
In this model, $X_{i}, Y_{i}$ are $2N \times 2N$ random Hermitian matrices with independent normally distributed elements of variance $1/3$, $N \gg 1$.  Since we address the WPs in superconducting stuctures at zero energy, in distinction from \cite{Wilkinson1,Wilkinson2,Wilkinson3}, we choose these random matrices to obey BdG mirror symmetry of the spectrum (class C \cite{Altland}). Generally, $l_c$ is defined as $l_c^{-3} = \sqrt{{\rm det}\langle\langle v_i v_j \rangle \rangle}/\delta^3_S$, $v_i \equiv \partial E /\partial \phi_i$, $\delta_S$ is the mean level spacing at the corresponding energy. For the model in use, $l_c =\pi \sqrt{3/2N}$ conveniently does not depend on $\vec{\phi}$ so that the WP density is uniform. 
We search the positions of WPs by an iterative minimization of the energy of the closest to zero level. To make sure we find all the WPs, we repeat the iteration cycle starting it from a randomly chosen point in the parameter space. We have to do this a number of times that by a factor exceeds the expected number of points. The execution time of the algorithm thus scales as $N^{9/2}$ so we cannot access very large $N$ and work with $N=40-80$.
For the WP concentration, we compute
\begin{equation}
\label{eq:estimations}
N_w/V = (0.83 \pm 0.05) l_c^{-3}.
\end{equation}
This is lower than the concentration of the level crossings in GUE ensemble \cite{Wilkinson2} $(2/3)\sqrt{\pi} l_c^{-3} \approx 1.18 l_c^{-3}$. We reproduce this result searching for the crossings of 10th and 11th levels.
\begin{figure}[tb]
\begin{center}
\includegraphics[width=0.92\linewidth]{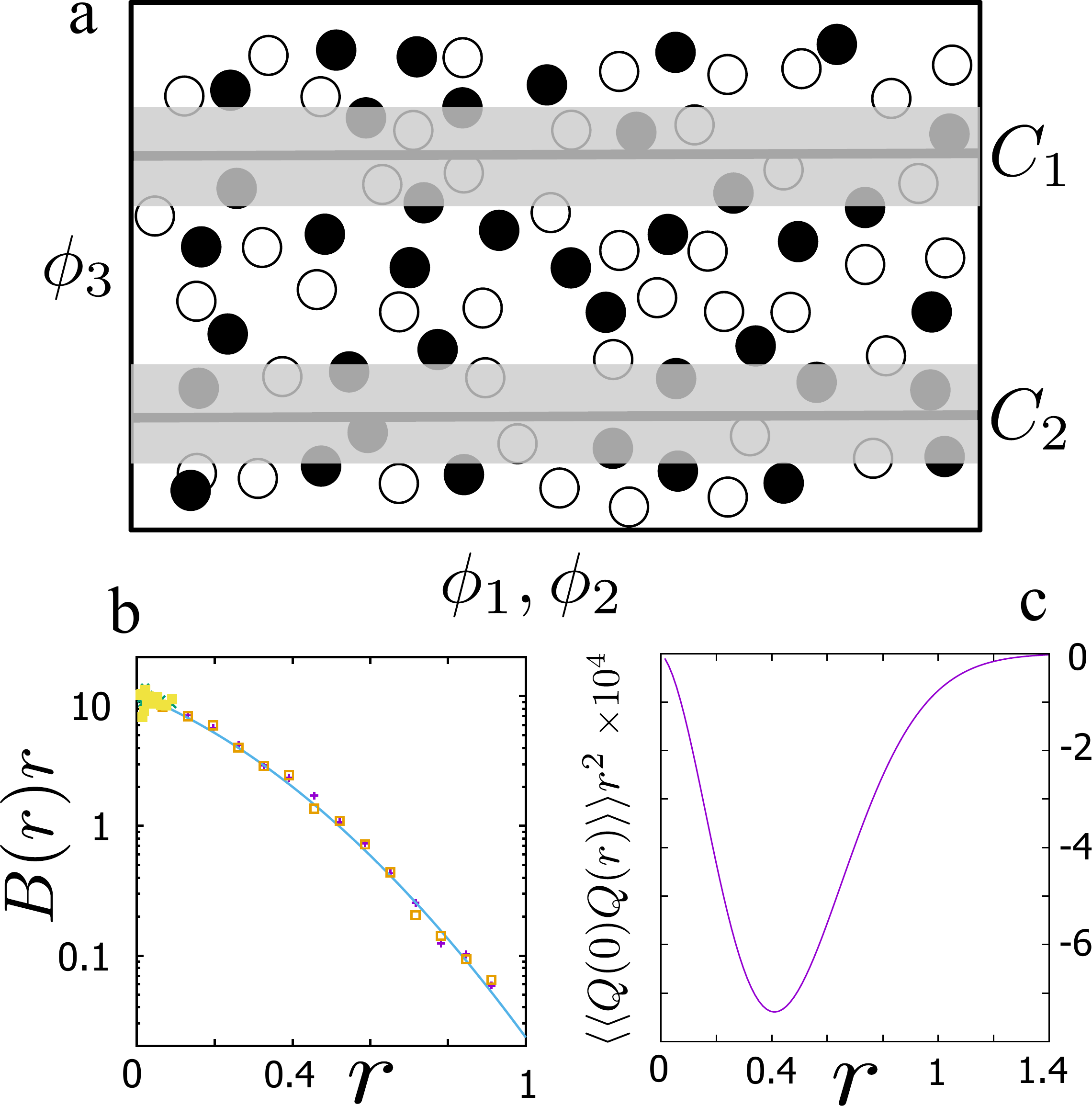}
\caption{Correlations of Weyl points. a. The distribution of WP charge is "  electroneutral". Owing to this, the fluctuations of Chern numbers in the planes $1,2$ are contributed by WPs at the distance $\simeq l_c$ from the planes (in grey strips). b. The numerical results for the correlator of Berry curvatures and the fit. c. The charge-charge correlator as computed from the fit.}
\label{fig:fig2}
\end{center}
\end{figure}

We address the correlator of charges of the WP's, $\langle\langle Q(0) Q(\vec{r})\rangle\rangle$, $\vec{r}$ being the vector distance in units of $l_c$. To enhance the statistics, we have evaluated an equivalent correlator of Berry curvatures of the closest to zero level. The results of $10^5$ runs per point are presented in Fig. \ref{fig:fig2} and can be fitted with 
\begin{equation}
\langle B^{\alpha}(0) B^{\beta}(\vec{r}) \rangle = \delta_{\alpha\beta} B(r), r B(r)\approx 10.4 e^{-2.8 r - 3.3 r^2}  
\end{equation}
Since the charge density of WPs is given by the divergence of Berry curvature (\cite{Berry2020,NewWilkinson}),
\begin{equation}
\langle\langle Q(0) Q(\vec{r})\rangle\rangle = (4\pi)^{-2} \nabla^2 B(r),
\end{equation}
see Fig. \ref{fig:fig2} c for the plot. By virtue of electro-neutrality of WP gas, $\int d\vec{r} \langle\langle Q(0) Q(\vec{r})\rangle\rangle = -N_w/V$. The fluctuations of Chern number over a surface of the size $\gg l_c$ are governed by $D \equiv -  \int d\vec{r} r \langle\langle Q(0) Q(\vec{r})\rangle\rangle$,
\begin{equation}
\langle\langle C^2 \rangle\rangle = D \int dS l_c^{-2}(\vec{\phi}) , 
\end{equation} 
$D \approx 0.5$ from our calculations, $d S$ being an area element of the surface. 
\begin{figure}[tb]
\begin{center}
\includegraphics[width=0.92\linewidth]{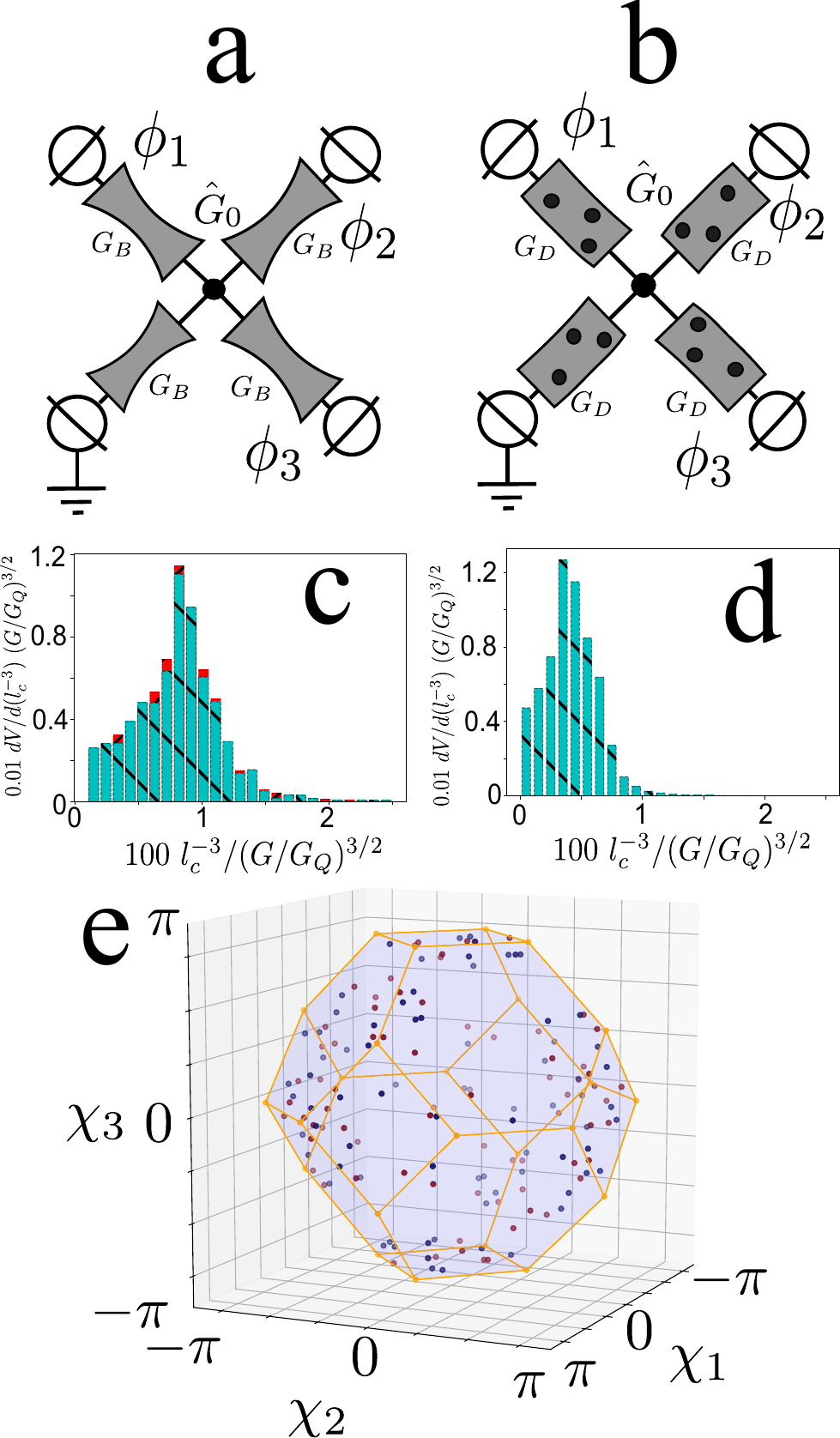}
\caption{Weyl points in concrete nanostructures. Example circuits: ballistic (a) and diffusive (b) crosses of identical arm conductances $G$. The results for $dV/d(l_c^-3)$ for the ballistic (c) and diffusive (d) cross. In (c), we compare estimations obtained from the analytical formula (green bars) and the actual positions of the WPs found (red bars) to demonstrate the correspondence within the statistical error. An example of WP positions found (e), $G/G_Q = 50$.    }
\label{fig:fig3}
\end{center}
\end{figure}

{\it Activity B.} While there are no perturbative methods to compute the density of WPs directly, they are available for the mesoscopic parametric correlations \cite{par1,par2}. With those, one can compute $l_c^{-3}$ for any system characterized by electronic Green functions. We make use of the quantum circuit theory \cite{QuantumTransport} that is a powerful finite-element technique for electronic Green functions. In quantum circuit theory, the structure is subdivided into reservoirs and nodes, the network is formed by connectors of various kinds, for instance, ballistic, tunnel or  diffusive. The Green functions are presented by the matrices $\hat{G}_a$, $\hat{G}^2 =1$, ${\rm Tr} \hat{G}=0$ defined in the nodes and and the reservoirs. The semiclassical solution is obtained by minimization of an action with respect to $\hat{G}$ in the reservoirs at fixed $\hat{G}$ in the nodes.

The mesoscopic parametric correlations for a general circuit theory have been derived in \cite{Campagnano}. For this, one substitutes to the action  $\hat{G}$ of double dimension, two diagonal blocks corresponding to the parameter sets $1,2$. Near the minimum, the action can be expanded up to quadratic terms with respect to non-diagonal deviations of $\check{G}$, $\check{M}$ being the matrix characterizing the quadratic expansion. The correlator of mesoscopic fluctuations of the action values at two parameter sets reads \cite{Campagnano}
\begin{equation}
\label{eq:Gqcorr}
\langle\langle {\cal S}_1 {\cal S}_2\rangle\rangle = \ln {\det}'{\check{M}}
\end{equation}  
where 'prime' excludes the zero eigenvalues of $\check{M}$ from the determinant.    

We implement this general technique to compute $l_c^{-3}$ for concrete superconducting nanostructures. It is known \cite{Beenakker} that the energies of Andreev bound states are expressed in terms of the effective scattering matrix $SS^*$, $S$ being the electron scattering matrix in the space of all channels coming to the nanostructure, $S^*$ being the hole scattering matrix,the superconducting phases included.  The circuit-theory action at the imaginary energy $\Delta \sin\theta$ (see e.g. \cite{Repin}) before the averaging over the mesoscopic fluctuations can be expressed in terms of eigenvalues $SS^* \to -e^{i\lambda}$ of the effective scattering matrix, these eigenvalues coming in $\pm$ pairs
\begin{align}
{\cal S}(\theta, \vec{\phi}) = -\sum_{\lambda} {\rm ln} \left( 1-\cos^2\theta\cos^2(\lambda/2)\right) \\
 \approx - \sum_{\lambda}{\rm ln} \left(\theta + i \lambda \right)
\end{align} 
the last equality holding for close to zero energies/eigenvalues. The correlator of the velocities in this limit is related to the correlator of the action values,
\begin{equation}
\langle\langle \partial_\alpha {\cal S}(\theta) \partial_\beta {\cal S}(\theta') \rangle \rangle  = \pi \frac{\langle\langle v_\alpha v_\beta  \rangle \rangle \rho_\lambda}{|\theta|+|\theta'|}.
\end{equation}
The action can be represented in a quantum circuit theory of $2\times 2$ matrices, and the correlator is to be computed with the aid of Eq. \ref{eq:Gqcorr}. In Supplementary Material \cite{supplmat}, we derive an explicit expression of $l_c^{-3}$ for a single-node structure. 

We concentrate on two simplest example MTSN (Fig. \ref{fig:fig3} a,b): a chaotic cavity connected to four leads by ballistic conductors of the same conductance $G$, ballistic cross, and the corresponding diffusive structure, diffusive cross. With the expressions obtained, we compute the distribution of $l_c^{-3}$, and, consequently, the WP density, over the phase space, by evaluating $l_c^{-3}$ in random points and collecting the data into histograms: this gives a fraction of the phase space volume $dV/d(l_c^{-3})$, at a given $l_c^{-3}$. The histograms for these two examples are qualitatively similar but distinct. Summing up the histograms and employing the result (\ref{eq:density}) gives the already mentioned estimations of the number of WP's, Eq. \ref{eq:estimations}.  
 
{\it Activity C.} We explicitly compute the WP positions for random chaotic cavities. For this, we pick up the electron scattering matrix $S$ from the circular orthogonal ensemble, and find all phase settings at which $S S^*$ has an eigenvalue $-1$ \cite{Beenakker}. We find 75-95 WPs for $N=G/G_Q=50$ conform to the results of the activities A,B and verify the scaling of the number of points with $N$. We also perform a more thorough check evaluating $l_c^{-3}$ in the random positions found and collecting the data to the histogram. The resulting estimation of $dV/d(l_c^{-3})$ that involves 2686 WP's coincides with the results of activity B within the statistical error (Fig. \ref{fig:fig3} c). In Fig. \ref{fig:fig3} e, we plot the positions of WP's found for a realization of $S$ at $G/G_Q=50$. We choose the coordinate system in the space of phases to be consistent with the symmetry of the structure, 
\begin{equation}
\chi_1 = \frac{1}{2}(\phi_1 - \phi_2 - \phi_3),
\end{equation}
$\chi_{2,3}$ are defined by the above relation with cyclically permuted indexes.
In these coordinates, the BZ is the truncated octahedron, as for a fcc lattice. The gapped region is in the centre of the BZ, the gapless region encloses its boundary \cite{Akhmerov}. The special points where the gapless region becomes infinitesimally thin \cite{Xiaoli} are located in the centres of the squares and hexagons, and, as seen in the Figure, the WPs are mostly concentrated in the corners of the BZ.     

Let us shortly discuss the methods of experimental detection of WPs in MTSN. For sufficiently large level splitting $G \simeq G_Q$, the WPs can be found spectroscopically as the zeros of the lowest Andreev state. For $G \simeq G_Q$ where the level splitting is small not exceeding $k_B T$, the detection is more challenging. For this case, we envisage the following detection methods: i. (Telegraph) noise measurements of the inductive or Berry curvature response of the MTSN. These responses diverge for a single discrete state at WP position. While the averaging over the states with thermal Boltzmann weights cancels the divergence, it is manifested in the noise at the time scale of the order of the time of switching between the states. 
ii. Transconductance (noise) measurements. We predict a transconductance $\simeq \sqrt{G G_Q}$. While in the presence of thermal averaging this transconductance is not quantized, its value will exhibit fluctuations as a function of the control phase \cite{grenoble} that can be used for scanning the WP positions.
iii. Transport spectroscopy. If the MTSN is in a weak tunnel contact with a normal lead, the differential conductance of this tunnel junction exhibits low-voltage anomalies at the WP positions \cite{Chen}. There is also a WP signature persisting at high voltage bias \cite{Chen}.

The source code and the raw data can be found at \cite{Hristo2021Dec}.

\begin{acknowledgments} 
This project has received funding from the European
Research Council (ERC) under the European Union's
Horizon 2020 research and innovation programme (grant
agreement No. 694272).
\end{acknowledgments}

\bibliography{abundance-bib}

\begin{thebibliography}{48}%
\makeatletter
\providecommand \@ifxundefined [1]{%
 \@ifx{#1\undefined}
}%
\providecommand \@ifnum [1]{%
 \ifnum #1\expandafter \@firstoftwo
 \else \expandafter \@secondoftwo
 \fi
}%
\providecommand \@ifx [1]{%
 \ifx #1\expandafter \@firstoftwo
 \else \expandafter \@secondoftwo
 \fi
}%
\providecommand \natexlab [1]{#1}%
\providecommand \enquote  [1]{``#1''}%
\providecommand \bibnamefont  [1]{#1}%
\providecommand \bibfnamefont [1]{#1}%
\providecommand \citenamefont [1]{#1}%
\providecommand \href@noop [0]{\@secondoftwo}%
\providecommand \href [0]{\begingroup \@sanitize@url \@href}%
\providecommand \@href[1]{\@@startlink{#1}\@@href}%
\providecommand \@@href[1]{\endgroup#1\@@endlink}%
\providecommand \@sanitize@url [0]{\catcode `\\12\catcode `\$12\catcode
  `\&12\catcode `\#12\catcode `\^12\catcode `\_12\catcode `\%12\relax}%
\providecommand \@@startlink[1]{}%
\providecommand \@@endlink[0]{}%
\providecommand \url  [0]{\begingroup\@sanitize@url \@url }%
\providecommand \@url [1]{\endgroup\@href {#1}{\urlprefix }}%
\providecommand \urlprefix  [0]{URL }%
\providecommand \Eprint [0]{\href }%
\providecommand \doibase [0]{https://doi.org/}%
\providecommand \selectlanguage [0]{\@gobble}%
\providecommand \bibinfo  [0]{\@secondoftwo}%
\providecommand \bibfield  [0]{\@secondoftwo}%
\providecommand \translation [1]{[#1]}%
\providecommand \BibitemOpen [0]{}%
\providecommand \bibitemStop [0]{}%
\providecommand \bibitemNoStop [0]{.\EOS\space}%
\providecommand \EOS [0]{\spacefactor3000\relax}%
\providecommand \BibitemShut  [1]{\csname bibitem#1\endcsname}%
\let\auto@bib@innerbib\@empty
\bibitem [{\citenamefont {Zhang}\ and\ \citenamefont
  {Das~Sarma}(2021{\natexlab{a}})}]{gen1}%
  \BibitemOpen
  \bibfield  {author} {\bibinfo {author} {\bibfnamefont {R.-X.}\ \bibnamefont
  {Zhang}}\ and\ \bibinfo {author} {\bibfnamefont {S.}~\bibnamefont
  {Das~Sarma}},\ }\bibfield  {title} {\bibinfo {title} {Anomalous floquet
  chiral topological superconductivity in a topological insulator sandwich
  structure},\ }\href {https://doi.org/10.1103/PhysRevLett.127.067001}
  {\bibfield  {journal} {\bibinfo  {journal} {Phys. Rev. Lett.}\ }\textbf
  {\bibinfo {volume} {127}},\ \bibinfo {pages} {067001} (\bibinfo {year}
  {2021}{\natexlab{a}})}\BibitemShut {NoStop}%
\bibitem [{\citenamefont {Zhao}\ \emph {et~al.}(2021)\citenamefont {Zhao},
  \citenamefont {Chen}, \citenamefont {Sheng},\ and\ \citenamefont
  {Yang}}]{gen2}%
  \BibitemOpen
  \bibfield  {author} {\bibinfo {author} {\bibfnamefont {Y.~X.}\ \bibnamefont
  {Zhao}}, \bibinfo {author} {\bibfnamefont {C.}~\bibnamefont {Chen}}, \bibinfo
  {author} {\bibfnamefont {X.-L.}\ \bibnamefont {Sheng}},\ and\ \bibinfo
  {author} {\bibfnamefont {S.~A.}\ \bibnamefont {Yang}},\ }\bibfield  {title}
  {\bibinfo {title} {Switching spinless and spinful topological phases with
  projective $pt$ symmetry},\ }\href
  {https://doi.org/10.1103/PhysRevLett.126.196402} {\bibfield  {journal}
  {\bibinfo  {journal} {Phys. Rev. Lett.}\ }\textbf {\bibinfo {volume} {126}},\
  \bibinfo {pages} {196402} (\bibinfo {year} {2021})}\BibitemShut {NoStop}%
\bibitem [{\citenamefont {Yang}\ \emph {et~al.}(2021)\citenamefont {Yang},
  \citenamefont {Yang}, \citenamefont {Hu},\ and\ \citenamefont {Liu}}]{gen3}%
  \BibitemOpen
  \bibfield  {author} {\bibinfo {author} {\bibfnamefont {Z.}~\bibnamefont
  {Yang}}, \bibinfo {author} {\bibfnamefont {Q.}~\bibnamefont {Yang}}, \bibinfo
  {author} {\bibfnamefont {J.}~\bibnamefont {Hu}},\ and\ \bibinfo {author}
  {\bibfnamefont {D.~E.}\ \bibnamefont {Liu}},\ }\bibfield  {title} {\bibinfo
  {title} {Dissipative floquet majorana modes in proximity-induced topological
  superconductors},\ }\href {https://doi.org/10.1103/PhysRevLett.126.086801}
  {\bibfield  {journal} {\bibinfo  {journal} {Phys. Rev. Lett.}\ }\textbf
  {\bibinfo {volume} {126}},\ \bibinfo {pages} {086801} (\bibinfo {year}
  {2021})}\BibitemShut {NoStop}%
\bibitem [{\citenamefont {Zhang}\ and\ \citenamefont
  {Das~Sarma}(2021{\natexlab{b}})}]{gen4}%
  \BibitemOpen
  \bibfield  {author} {\bibinfo {author} {\bibfnamefont {R.-X.}\ \bibnamefont
  {Zhang}}\ and\ \bibinfo {author} {\bibfnamefont {S.}~\bibnamefont
  {Das~Sarma}},\ }\bibfield  {title} {\bibinfo {title} {Intrinsic
  time-reversal-invariant topological superconductivity in thin films of
  iron-based superconductors},\ }\href
  {https://doi.org/10.1103/PhysRevLett.126.137001} {\bibfield  {journal}
  {\bibinfo  {journal} {Phys. Rev. Lett.}\ }\textbf {\bibinfo {volume} {126}},\
  \bibinfo {pages} {137001} (\bibinfo {year} {2021}{\natexlab{b}})}\BibitemShut
  {NoStop}%
\bibitem [{\citenamefont {Hasan}\ and\ \citenamefont
  {Kane}(2010)}]{Topinsulators}%
  \BibitemOpen
  \bibfield  {author} {\bibinfo {author} {\bibfnamefont {M.~Z.}\ \bibnamefont
  {Hasan}}\ and\ \bibinfo {author} {\bibfnamefont {C.~L.}\ \bibnamefont
  {Kane}},\ }\bibfield  {title} {\bibinfo {title} {Colloquium: Topological
  insulators},\ }\href {https://doi.org/10.1103/RevModPhys.82.3045} {\bibfield
  {journal} {\bibinfo  {journal} {Rev. Mod. Phys.}\ }\textbf {\bibinfo {volume}
  {82}},\ \bibinfo {pages} {3045} (\bibinfo {year} {2010})}\BibitemShut
  {NoStop}%
\bibitem [{\citenamefont {Qi}\ and\ \citenamefont
  {Zhang}(2011)}]{TopSuperconductors}%
  \BibitemOpen
  \bibfield  {author} {\bibinfo {author} {\bibfnamefont {X.-L.}\ \bibnamefont
  {Qi}}\ and\ \bibinfo {author} {\bibfnamefont {S.-C.}\ \bibnamefont {Zhang}},\
  }\bibfield  {title} {\bibinfo {title} {Topological insulators and
  superconductors},\ }\href {https://doi.org/10.1103/RevModPhys.83.1057}
  {\bibfield  {journal} {\bibinfo  {journal} {Rev. Mod. Phys.}\ }\textbf
  {\bibinfo {volume} {83}},\ \bibinfo {pages} {1057} (\bibinfo {year}
  {2011})}\BibitemShut {NoStop}%
\bibitem [{\citenamefont {Bernevig}\ and\ \citenamefont
  {Hughes}(2013)}]{bernevig2013topological}%
  \BibitemOpen
  \bibfield  {author} {\bibinfo {author} {\bibfnamefont {B.}~\bibnamefont
  {Bernevig}}\ and\ \bibinfo {author} {\bibfnamefont {T.}~\bibnamefont
  {Hughes}},\ }\href {https://books.google.nl/books?id=wOn7JHSSxrsC} {\emph
  {\bibinfo {title} {Topological Insulators and Topological Superconductors}}}\
  (\bibinfo  {publisher} {Princeton University Press},\ \bibinfo {year}
  {2013})\BibitemShut {NoStop}%
\bibitem [{\citenamefont {Weyl}(1929)}]{Weyl}%
  \BibitemOpen
  \bibfield  {author} {\bibinfo {author} {\bibfnamefont {H.}~\bibnamefont
  {Weyl}},\ }\bibfield  {title} {\bibinfo {title} {Electron und
  {Gravitation}},\ }\href@noop {} {\bibfield  {journal} {\bibinfo  {journal}
  {Z.\ Phys.}\ }\textbf {\bibinfo {volume} {56}},\ \bibinfo {pages} {330}
  (\bibinfo {year} {1929})}\BibitemShut {NoStop}%
\bibitem [{\citenamefont {Armitage}\ \emph {et~al.}(2018)\citenamefont
  {Armitage}, \citenamefont {Mele},\ and\ \citenamefont
  {Vishwanath}}]{WeylSemimetals}%
  \BibitemOpen
  \bibfield  {author} {\bibinfo {author} {\bibfnamefont {N.~P.}\ \bibnamefont
  {Armitage}}, \bibinfo {author} {\bibfnamefont {E.~J.}\ \bibnamefont {Mele}},\
  and\ \bibinfo {author} {\bibfnamefont {A.}~\bibnamefont {Vishwanath}},\
  }\bibfield  {title} {\bibinfo {title} {Weyl and dirac semimetals in
  three-dimensional solids},\ }\href
  {https://doi.org/10.1103/RevModPhys.90.015001} {\bibfield  {journal}
  {\bibinfo  {journal} {Rev. Mod. Phys.}\ }\textbf {\bibinfo {volume} {90}},\
  \bibinfo {pages} {015001} (\bibinfo {year} {2018})}\BibitemShut {NoStop}%
\bibitem [{\citenamefont {Faure}\ and\ \citenamefont
  {Zhilinskii}(2000)}]{faure}%
  \BibitemOpen
  \bibfield  {author} {\bibinfo {author} {\bibfnamefont {F.}~\bibnamefont
  {Faure}}\ and\ \bibinfo {author} {\bibfnamefont {B.~I.}\ \bibnamefont
  {Zhilinskii}},\ }\bibfield  {title} {\bibinfo {title} {Topological {Chern}
  indices in molecular spectra},\ }\href@noop {} {\bibfield  {journal}
  {\bibinfo  {journal} {Phys.\ Rev.\ Lett.}\ }\textbf {\bibinfo {volume}
  {85}},\ \bibinfo {pages} {960} (\bibinfo {year} {2000})}\BibitemShut
  {NoStop}%
\bibitem [{\citenamefont {Wernsdorfer}\ \emph {et~al.}(2005)\citenamefont
  {Wernsdorfer}, \citenamefont {Chakov},\ and\ \citenamefont
  {G.~Christou}}]{werns2}%
  \BibitemOpen
  \bibfield  {author} {\bibinfo {author} {\bibfnamefont {W.}~\bibnamefont
  {Wernsdorfer}}, \bibinfo {author} {\bibfnamefont {N.~E.}\ \bibnamefont
  {Chakov}},\ and\ \bibinfo {author} {\bibfnamefont {G.}~\bibnamefont
  {G.~Christou}},\ }\bibfield  {title} {\bibinfo {title} {Quantum phase
  interference and parity effects in {Mn${}_12$} single-molecule magnets},\
  }\href@noop {} {\bibfield  {journal} {\bibinfo  {journal} {Phys.\ Rev.\
  Lett.}\ }\textbf {\bibinfo {volume} {95}},\ \bibinfo {pages} {037203}
  (\bibinfo {year} {2005})}\BibitemShut {NoStop}%
\bibitem [{\citenamefont {Leone}\ \emph {et~al.}(2008)\citenamefont {Leone},
  \citenamefont {L{\'e}vy},\ and\ \citenamefont {Lafarge}}]{leone2008cooper}%
  \BibitemOpen
  \bibfield  {author} {\bibinfo {author} {\bibfnamefont {R.}~\bibnamefont
  {Leone}}, \bibinfo {author} {\bibfnamefont {L.~P.}\ \bibnamefont
  {L{\'e}vy}},\ and\ \bibinfo {author} {\bibfnamefont {P.}~\bibnamefont
  {Lafarge}},\ }\bibfield  {title} {\bibinfo {title} {Cooper-pair pump as a
  quantized current source},\ }\href@noop {} {\bibfield  {journal} {\bibinfo
  {journal} {Phys.\ Rev.\ Lett.}\ }\textbf {\bibinfo {volume} {100}},\ \bibinfo
  {pages} {117001} (\bibinfo {year} {2008})}\BibitemShut {NoStop}%
\bibitem [{\citenamefont {Riwar}\ \emph {et~al.}(2016)\citenamefont {Riwar},
  \citenamefont {Houzet}, \citenamefont {Meyer},\ and\ \citenamefont
  {Nazarov}}]{grenoble}%
  \BibitemOpen
  \bibfield  {author} {\bibinfo {author} {\bibfnamefont {{\relax
  R.-P}.}~\bibnamefont {Riwar}}, \bibinfo {author} {\bibfnamefont
  {M.}~\bibnamefont {Houzet}}, \bibinfo {author} {\bibfnamefont {J.~S.}\
  \bibnamefont {Meyer}},\ and\ \bibinfo {author} {\bibfnamefont {Y.~V.}\
  \bibnamefont {Nazarov}},\ }\bibfield  {title} {\bibinfo {title}
  {Multi-terminal {Josephson} junctions as topological matter},\ }\href@noop {}
  {\bibfield  {journal} {\bibinfo  {journal} {Nat.\ Comm.}\ }\textbf {\bibinfo
  {volume} {7}},\ \bibinfo {pages} {11167} (\bibinfo {year}
  {2016})}\BibitemShut {NoStop}%
\bibitem [{\citenamefont {Xie}\ and\ \citenamefont {Levchenko}(2019)}]{thy1}%
  \BibitemOpen
  \bibfield  {author} {\bibinfo {author} {\bibfnamefont {H.-Y.}\ \bibnamefont
  {Xie}}\ and\ \bibinfo {author} {\bibfnamefont {A.}~\bibnamefont
  {Levchenko}},\ }\bibfield  {title} {\bibinfo {title} {Topological
  supercurrents interaction and fluctuations in the multiterminal josephson
  effect},\ }\href {https://doi.org/10.1103/PhysRevB.99.094519} {\bibfield
  {journal} {\bibinfo  {journal} {Phys. Rev. B}\ }\textbf {\bibinfo {volume}
  {99}},\ \bibinfo {pages} {094519} (\bibinfo {year} {2019})}\BibitemShut
  {NoStop}%
\bibitem [{\citenamefont {Xie}\ \emph {et~al.}(2017)\citenamefont {Xie},
  \citenamefont {Vavilov},\ and\ \citenamefont {Levchenko}}]{thy2}%
  \BibitemOpen
  \bibfield  {author} {\bibinfo {author} {\bibfnamefont {H.-Y.}\ \bibnamefont
  {Xie}}, \bibinfo {author} {\bibfnamefont {M.~G.}\ \bibnamefont {Vavilov}},\
  and\ \bibinfo {author} {\bibfnamefont {A.}~\bibnamefont {Levchenko}},\
  }\bibfield  {title} {\bibinfo {title} {Topological andreev bands in
  three-terminal josephson junctions},\ }\href
  {https://doi.org/10.1103/PhysRevB.96.161406} {\bibfield  {journal} {\bibinfo
  {journal} {Phys. Rev. B}\ }\textbf {\bibinfo {volume} {96}},\ \bibinfo
  {pages} {161406} (\bibinfo {year} {2017})}\BibitemShut {NoStop}%
\bibitem [{\citenamefont {Houzet}\ and\ \citenamefont {Meyer}(2019)}]{thy3}%
  \BibitemOpen
  \bibfield  {author} {\bibinfo {author} {\bibfnamefont {M.}~\bibnamefont
  {Houzet}}\ and\ \bibinfo {author} {\bibfnamefont {J.~S.}\ \bibnamefont
  {Meyer}},\ }\bibfield  {title} {\bibinfo {title} {{Majorana-Weyl crossings in
  topological multiterminal junctions}},\ }\bibfield  {journal} {\bibinfo
  {journal} {{Phys. Rev. B}}\ }\textbf {\bibinfo {volume} {{100}}},\ \href
  {https://doi.org/{10.1103/PhysRevB.100.014521}}
  {{10.1103/PhysRevB.100.014521}} (\bibinfo {year} {{2019}})\BibitemShut
  {NoStop}%
\bibitem [{\citenamefont {Klees}\ \emph {et~al.}(2020)\citenamefont {Klees},
  \citenamefont {Rastelli}, \citenamefont {Cuevas},\ and\ \citenamefont
  {Belzig}}]{thy4}%
  \BibitemOpen
  \bibfield  {author} {\bibinfo {author} {\bibfnamefont {R.~L.}\ \bibnamefont
  {Klees}}, \bibinfo {author} {\bibfnamefont {G.}~\bibnamefont {Rastelli}},
  \bibinfo {author} {\bibfnamefont {J.~C.}\ \bibnamefont {Cuevas}},\ and\
  \bibinfo {author} {\bibfnamefont {W.}~\bibnamefont {Belzig}},\ }\bibfield
  {title} {\bibinfo {title} {{Microwave Spectroscopy Reveals the Quantum
  Geometric Tensor of Topological Josephson Matter}},\ }\bibfield  {journal}
  {\bibinfo  {journal} {{Phys. Rev. Lett.}}\ }\textbf {\bibinfo {volume}
  {{124}}},\ \href {https://doi.org/{10.1103/PhysRevLett.124.197002}}
  {{10.1103/PhysRevLett.124.197002}} (\bibinfo {year} {{2020}})\BibitemShut
  {NoStop}%
\bibitem [{\citenamefont {Fatemi}\ \emph {et~al.}(2021)\citenamefont {Fatemi},
  \citenamefont {Akhmerov},\ and\ \citenamefont {Bretheau}}]{thy5}%
  \BibitemOpen
  \bibfield  {author} {\bibinfo {author} {\bibfnamefont {V.}~\bibnamefont
  {Fatemi}}, \bibinfo {author} {\bibfnamefont {A.~R.}\ \bibnamefont
  {Akhmerov}},\ and\ \bibinfo {author} {\bibfnamefont {L.}~\bibnamefont
  {Bretheau}},\ }\bibfield  {title} {\bibinfo {title} {Weyl josephson
  circuits},\ }\href {https://doi.org/10.1103/PhysRevResearch.3.013288}
  {\bibfield  {journal} {\bibinfo  {journal} {Phys. Rev. Research}\ }\textbf
  {\bibinfo {volume} {3}},\ \bibinfo {pages} {013288} (\bibinfo {year}
  {2021})}\BibitemShut {NoStop}%
\bibitem [{\citenamefont {Weisbrich}\ \emph {et~al.}(2021)\citenamefont
  {Weisbrich}, \citenamefont {Rastelli},\ and\ \citenamefont {Belzig}}]{thy6}%
  \BibitemOpen
  \bibfield  {author} {\bibinfo {author} {\bibfnamefont {H.}~\bibnamefont
  {Weisbrich}}, \bibinfo {author} {\bibfnamefont {G.}~\bibnamefont
  {Rastelli}},\ and\ \bibinfo {author} {\bibfnamefont {W.}~\bibnamefont
  {Belzig}},\ }\bibfield  {title} {\bibinfo {title} {{Geometrical Rabi
  oscillations and Landau-Zener transitions in non-Abelian systems}},\
  }\bibfield  {journal} {\bibinfo  {journal} {{Phys. Rev. Res.}}\ }\textbf
  {\bibinfo {volume} {{3}}},\ \href
  {https://doi.org/{10.1103/PhysRevResearch.3.033122}}
  {{10.1103/PhysRevResearch.3.033122}} (\bibinfo {year} {{2021}})\BibitemShut
  {NoStop}%
\bibitem [{\citenamefont {Yokoyama}\ and\ \citenamefont
  {Nazarov}(2015)}]{thy7}%
  \BibitemOpen
  \bibfield  {author} {\bibinfo {author} {\bibfnamefont {T.}~\bibnamefont
  {Yokoyama}}\ and\ \bibinfo {author} {\bibfnamefont {Y.~V.}\ \bibnamefont
  {Nazarov}},\ }\bibfield  {title} {\bibinfo {title} {Singularities in the
  andreev spectrum of a multiterminal josephson junction},\ }\href
  {https://doi.org/10.1103/physrevb.92.155437} {\bibfield  {journal} {\bibinfo
  {journal} {Physical Review B}\ }\textbf {\bibinfo {volume} {92}},\ \bibinfo
  {pages} {155437} (\bibinfo {year} {2015})}\BibitemShut {NoStop}%
\bibitem [{\citenamefont {Pankratova}\ \emph {et~al.}(2020)\citenamefont
  {Pankratova}, \citenamefont {Lee}, \citenamefont {Kuzmin}, \citenamefont
  {Wickramasinghe}, \citenamefont {Mayer}, \citenamefont {Yuan}, \citenamefont
  {Vavilov}, \citenamefont {Shabani},\ and\ \citenamefont
  {Manucharyan}}]{exp1}%
  \BibitemOpen
  \bibfield  {author} {\bibinfo {author} {\bibfnamefont {N.}~\bibnamefont
  {Pankratova}}, \bibinfo {author} {\bibfnamefont {H.}~\bibnamefont {Lee}},
  \bibinfo {author} {\bibfnamefont {R.}~\bibnamefont {Kuzmin}}, \bibinfo
  {author} {\bibfnamefont {K.}~\bibnamefont {Wickramasinghe}}, \bibinfo
  {author} {\bibfnamefont {W.}~\bibnamefont {Mayer}}, \bibinfo {author}
  {\bibfnamefont {J.}~\bibnamefont {Yuan}}, \bibinfo {author} {\bibfnamefont
  {M.~G.}\ \bibnamefont {Vavilov}}, \bibinfo {author} {\bibfnamefont
  {J.}~\bibnamefont {Shabani}},\ and\ \bibinfo {author} {\bibfnamefont {V.~E.}\
  \bibnamefont {Manucharyan}},\ }\bibfield  {title} {\bibinfo {title}
  {{Multiterminal Josephson Effect}},\ }\bibfield  {journal} {\bibinfo
  {journal} {{Phys.\ Rev.\ X}}\ }\textbf {\bibinfo {volume} {{10}}},\ \href
  {https://doi.org/{10.1103/PhysRevX.10.031051}} {{10.1103/PhysRevX.10.031051}}
  (\bibinfo {year} {{2020}})\BibitemShut {NoStop}%
\bibitem [{\citenamefont {Graziano}\ \emph {et~al.}(2020)\citenamefont
  {Graziano}, \citenamefont {Lee}, \citenamefont {Pendharkar}, \citenamefont
  {Palmstrom},\ and\ \citenamefont {Pribiag}}]{exp2}%
  \BibitemOpen
  \bibfield  {author} {\bibinfo {author} {\bibfnamefont {G.}~\bibnamefont
  {Graziano}, \bibfnamefont {V}}, \bibinfo {author} {\bibfnamefont {J.~S.}\
  \bibnamefont {Lee}}, \bibinfo {author} {\bibfnamefont {M.}~\bibnamefont
  {Pendharkar}}, \bibinfo {author} {\bibfnamefont {C.}~\bibnamefont
  {Palmstrom}},\ and\ \bibinfo {author} {\bibfnamefont {V.~S.}\ \bibnamefont
  {Pribiag}},\ }\bibfield  {title} {\bibinfo {title} {{Transport studies in a
  gate-tunable three-terminal Josephson junction}},\ }\bibfield  {journal}
  {\bibinfo  {journal} {{Phys. Rev. B}}\ }\textbf {\bibinfo {volume} {{101}}},\
  \href {https://doi.org/{10.1103/PhysRevB.101.054510}}
  {{10.1103/PhysRevB.101.054510}} (\bibinfo {year} {{2020}})\BibitemShut
  {NoStop}%
\bibitem [{\citenamefont {Draelos}\ \emph {et~al.}(2019)\citenamefont
  {Draelos}, \citenamefont {Wei}, \citenamefont {Seredinski}, \citenamefont
  {Li}, \citenamefont {Mehta}, \citenamefont {Watanabe}, \citenamefont
  {Taniguchi}, \citenamefont {Borzenets}, \citenamefont {Amet},\ and\
  \citenamefont {Finkelstein}}]{exp3}%
  \BibitemOpen
  \bibfield  {author} {\bibinfo {author} {\bibfnamefont {A.~W.}\ \bibnamefont
  {Draelos}}, \bibinfo {author} {\bibfnamefont {M.-T.}\ \bibnamefont {Wei}},
  \bibinfo {author} {\bibfnamefont {A.}~\bibnamefont {Seredinski}}, \bibinfo
  {author} {\bibfnamefont {H.}~\bibnamefont {Li}}, \bibinfo {author}
  {\bibfnamefont {Y.}~\bibnamefont {Mehta}}, \bibinfo {author} {\bibfnamefont
  {K.}~\bibnamefont {Watanabe}}, \bibinfo {author} {\bibfnamefont
  {T.}~\bibnamefont {Taniguchi}}, \bibinfo {author} {\bibfnamefont {I.~V.}\
  \bibnamefont {Borzenets}}, \bibinfo {author} {\bibfnamefont {F.}~\bibnamefont
  {Amet}},\ and\ \bibinfo {author} {\bibfnamefont {G.}~\bibnamefont
  {Finkelstein}},\ }\bibfield  {title} {\bibinfo {title} {{Supercurrent Flow in
  Multiterminal Graphene Josephson Junctions}},\ }\href
  {https://doi.org/{10.1021/acs.nanolett.8b04330}} {\bibfield  {journal}
  {\bibinfo  {journal} {{Nano\ Lett.}}\ }\textbf {\bibinfo {volume} {{19}}},\
  \bibinfo {pages} {1039} (\bibinfo {year} {{2019}})}\BibitemShut {NoStop}%
\bibitem [{\citenamefont {Pfeffer}\ \emph {et~al.}(2014)\citenamefont
  {Pfeffer}, \citenamefont {Duvauchelle}, \citenamefont {Courtois},
  \citenamefont {M\'elin}, \citenamefont {Feinberg},\ and\ \citenamefont
  {Lefloch}}]{exp4}%
  \BibitemOpen
  \bibfield  {author} {\bibinfo {author} {\bibfnamefont {A.~H.}\ \bibnamefont
  {Pfeffer}}, \bibinfo {author} {\bibfnamefont {J.~E.}\ \bibnamefont
  {Duvauchelle}}, \bibinfo {author} {\bibfnamefont {H.}~\bibnamefont
  {Courtois}}, \bibinfo {author} {\bibfnamefont {R.}~\bibnamefont {M\'elin}},
  \bibinfo {author} {\bibfnamefont {D.}~\bibnamefont {Feinberg}},\ and\
  \bibinfo {author} {\bibfnamefont {F.}~\bibnamefont {Lefloch}},\ }\bibfield
  {title} {\bibinfo {title} {Subgap structure in the conductance of a
  three-terminal josephson junction},\ }\href
  {https://doi.org/10.1103/PhysRevB.90.075401} {\bibfield  {journal} {\bibinfo
  {journal} {Phys. Rev. B}\ }\textbf {\bibinfo {volume} {90}},\ \bibinfo
  {pages} {075401} (\bibinfo {year} {2014})}\BibitemShut {NoStop}%
\bibitem [{\citenamefont {Fadaly}\ \emph {et~al.}(2017)\citenamefont {Fadaly},
  \citenamefont {Zhang}, \citenamefont {Conesa-Boj}, \citenamefont {Car},
  \citenamefont {G{\ifmmode\ddot{u}\else\"{u}\fi}l}, \citenamefont {Plissard},
  \citenamefont {Op~het Veld}, \citenamefont
  {K{\ifmmode\ddot{o}\else\"{o}\fi}lling}, \citenamefont {Kouwenhoven},\ and\
  \citenamefont {Bakkers}}]{exp5}%
  \BibitemOpen
  \bibfield  {author} {\bibinfo {author} {\bibfnamefont {E.~M.~T.}\
  \bibnamefont {Fadaly}}, \bibinfo {author} {\bibfnamefont {H.}~\bibnamefont
  {Zhang}}, \bibinfo {author} {\bibfnamefont {S.}~\bibnamefont {Conesa-Boj}},
  \bibinfo {author} {\bibfnamefont {D.}~\bibnamefont {Car}}, \bibinfo {author}
  {\bibfnamefont {{\ifmmode\ddot{O}\else\"{O}\fi}.}~\bibnamefont
  {G{\ifmmode\ddot{u}\else\"{u}\fi}l}}, \bibinfo {author} {\bibfnamefont
  {S.~R.}\ \bibnamefont {Plissard}}, \bibinfo {author} {\bibfnamefont
  {R.~L.~M.}\ \bibnamefont {Op~het Veld}}, \bibinfo {author} {\bibfnamefont
  {S.}~\bibnamefont {K{\ifmmode\ddot{o}\else\"{o}\fi}lling}}, \bibinfo {author}
  {\bibfnamefont {L.~P.}\ \bibnamefont {Kouwenhoven}},\ and\ \bibinfo {author}
  {\bibfnamefont {E.~P. A.~M.}\ \bibnamefont {Bakkers}},\ }\bibfield  {title}
  {\bibinfo {title} {{Observation of Conductance Quantization in InSb Nanowire
  Networks}},\ }\href {https://doi.org/10.1021/acs.nanolett.7b00797} {\bibfield
   {journal} {\bibinfo  {journal} {Nano Lett.}\ }\textbf {\bibinfo {volume}
  {17}},\ \bibinfo {pages} {6511} (\bibinfo {year} {2017})},\ \bibinfo {note}
  {pMID: 28665621},\ \Eprint
  {https://arxiv.org/abs/https://doi.org/10.1021/acs.nanolett.7b00797}
  {https://doi.org/10.1021/acs.nanolett.7b00797} \BibitemShut {NoStop}%
\bibitem [{\citenamefont {Padurariu}\ \emph {et~al.}(2015)\citenamefont
  {Padurariu}, \citenamefont {Jonckheere}, \citenamefont {Rech}, \citenamefont
  {M\'elin}, \citenamefont {Feinberg}, \citenamefont {Martin},\ and\
  \citenamefont {Nazarov}}]{Padurariu}%
  \BibitemOpen
  \bibfield  {author} {\bibinfo {author} {\bibfnamefont {C.}~\bibnamefont
  {Padurariu}}, \bibinfo {author} {\bibfnamefont {T.}~\bibnamefont
  {Jonckheere}}, \bibinfo {author} {\bibfnamefont {J.}~\bibnamefont {Rech}},
  \bibinfo {author} {\bibfnamefont {R.}~\bibnamefont {M\'elin}}, \bibinfo
  {author} {\bibfnamefont {D.}~\bibnamefont {Feinberg}}, \bibinfo {author}
  {\bibfnamefont {T.}~\bibnamefont {Martin}},\ and\ \bibinfo {author}
  {\bibfnamefont {Y.~V.}\ \bibnamefont {Nazarov}},\ }\bibfield  {title}
  {\bibinfo {title} {Closing the proximity gap in a metallic josephson junction
  between three superconductors},\ }\href
  {https://doi.org/10.1103/PhysRevB.92.205409} {\bibfield  {journal} {\bibinfo
  {journal} {Phys. Rev. B}\ }\textbf {\bibinfo {volume} {92}},\ \bibinfo
  {pages} {205409} (\bibinfo {year} {2015})}\BibitemShut {NoStop}%
\bibitem [{\citenamefont {Strambini}\ \emph {et~al.}(2016)\citenamefont
  {Strambini}, \citenamefont {D'Ambrosio}, \citenamefont {Vischi},
  \citenamefont {Bergeret}, \citenamefont {Nazarov},\ and\ \citenamefont
  {Giazotto}}]{omega1}%
  \BibitemOpen
  \bibfield  {author} {\bibinfo {author} {\bibfnamefont {E.}~\bibnamefont
  {Strambini}}, \bibinfo {author} {\bibfnamefont {S.}~\bibnamefont
  {D'Ambrosio}}, \bibinfo {author} {\bibfnamefont {F.}~\bibnamefont {Vischi}},
  \bibinfo {author} {\bibfnamefont {F.~S.}\ \bibnamefont {Bergeret}}, \bibinfo
  {author} {\bibfnamefont {Y.~V.}\ \bibnamefont {Nazarov}},\ and\ \bibinfo
  {author} {\bibfnamefont {F.}~\bibnamefont {Giazotto}},\ }\bibfield  {title}
  {\bibinfo {title} {The o-squipt as a tool to phase-engineer josephson
  topological materials},\ }\href {https://doi.org/10.1038/nnano.2016.157}
  {\bibfield  {journal} {\bibinfo  {journal} {Nature Nanotechnology}\ }\textbf
  {\bibinfo {volume} {11}},\ \bibinfo {pages} {1055 EP } (\bibinfo {year}
  {2016})}\BibitemShut {NoStop}%
\bibitem [{\citenamefont {Vischi}\ \emph {et~al.}(2017)\citenamefont {Vischi},
  \citenamefont {Carrega}, \citenamefont {Strambini}, \citenamefont
  {D'Ambrosio}, \citenamefont {Bergeret}, \citenamefont {Nazarov},\ and\
  \citenamefont {Giazotto}}]{omega2}%
  \BibitemOpen
  \bibfield  {author} {\bibinfo {author} {\bibfnamefont {F.}~\bibnamefont
  {Vischi}}, \bibinfo {author} {\bibfnamefont {M.}~\bibnamefont {Carrega}},
  \bibinfo {author} {\bibfnamefont {E.}~\bibnamefont {Strambini}}, \bibinfo
  {author} {\bibfnamefont {S.}~\bibnamefont {D'Ambrosio}}, \bibinfo {author}
  {\bibfnamefont {F.~S.}\ \bibnamefont {Bergeret}}, \bibinfo {author}
  {\bibfnamefont {Y.~V.}\ \bibnamefont {Nazarov}},\ and\ \bibinfo {author}
  {\bibfnamefont {F.}~\bibnamefont {Giazotto}},\ }\bibfield  {title} {\bibinfo
  {title} {Coherent transport properties of a three-terminal hybrid
  superconducting interferometer},\ }\href
  {https://doi.org/10.1103/PhysRevB.95.054504} {\bibfield  {journal} {\bibinfo
  {journal} {Phys. Rev. B}\ }\textbf {\bibinfo {volume} {95}},\ \bibinfo
  {pages} {054504} (\bibinfo {year} {2017})}\BibitemShut {NoStop}%
\bibitem [{\citenamefont {Yokoyama}\ \emph {et~al.}(2017)\citenamefont
  {Yokoyama}, \citenamefont {Reutlinger}, \citenamefont {Belzig},\ and\
  \citenamefont {Nazarov}}]{Yokoyama2017}%
  \BibitemOpen
  \bibfield  {author} {\bibinfo {author} {\bibfnamefont {T.}~\bibnamefont
  {Yokoyama}}, \bibinfo {author} {\bibfnamefont {J.}~\bibnamefont
  {Reutlinger}}, \bibinfo {author} {\bibfnamefont {W.}~\bibnamefont {Belzig}},\
  and\ \bibinfo {author} {\bibfnamefont {Y.~V.}\ \bibnamefont {Nazarov}},\
  }\bibfield  {title} {\bibinfo {title} {Order, disorder, and tunable gaps in
  the spectrum of andreev bound states in a multiterminal superconducting
  device},\ }\href {https://doi.org/10.1103/PhysRevB.95.045411} {\bibfield
  {journal} {\bibinfo  {journal} {Phys. Rev. B}\ }\textbf {\bibinfo {volume}
  {95}},\ \bibinfo {pages} {045411} (\bibinfo {year} {2017})}\BibitemShut
  {NoStop}%
\bibitem [{\citenamefont {Huang}\ and\ \citenamefont {Nazarov}(2019)}]{Xiaoli}%
  \BibitemOpen
  \bibfield  {author} {\bibinfo {author} {\bibfnamefont {X.-L.}\ \bibnamefont
  {Huang}}\ and\ \bibinfo {author} {\bibfnamefont {Y.~V.}\ \bibnamefont
  {Nazarov}},\ }\bibfield  {title} {\bibinfo {title} {Topology
  protection--unprotection transition: Example from multiterminal
  superconducting nanostructures},\ }\href
  {https://doi.org/10.1103/PhysRevB.100.085408} {\bibfield  {journal} {\bibinfo
   {journal} {Phys. Rev. B}\ }\textbf {\bibinfo {volume} {100}},\ \bibinfo
  {pages} {085408} (\bibinfo {year} {2019})}\BibitemShut {NoStop}%
\bibitem [{\citenamefont {Simons}\ and\ \citenamefont
  {Altshuler}(1993)}]{universal1}%
  \BibitemOpen
  \bibfield  {author} {\bibinfo {author} {\bibfnamefont {B.~D.}\ \bibnamefont
  {Simons}}\ and\ \bibinfo {author} {\bibfnamefont {B.~L.}\ \bibnamefont
  {Altshuler}},\ }\bibfield  {title} {\bibinfo {title} {Universal velocity
  correlations in disordered and chaotic systems},\ }\href
  {https://doi.org/10.1103/PhysRevLett.70.4063} {\bibfield  {journal} {\bibinfo
   {journal} {Phys. Rev. Lett.}\ }\textbf {\bibinfo {volume} {70}},\ \bibinfo
  {pages} {4063} (\bibinfo {year} {1993})}\BibitemShut {NoStop}%
\bibitem [{\citenamefont {Beenakker}(1993)}]{universal2}%
  \BibitemOpen
  \bibfield  {author} {\bibinfo {author} {\bibfnamefont {C.~W.~J.}\
  \bibnamefont {Beenakker}},\ }\bibfield  {title} {\bibinfo {title}
  {Brownian-motion model for parametric correlations in the spectra of
  disordered metals},\ }\href {https://doi.org/10.1103/PhysRevLett.70.4126}
  {\bibfield  {journal} {\bibinfo  {journal} {Phys. Rev. Lett.}\ }\textbf
  {\bibinfo {volume} {70}},\ \bibinfo {pages} {4126} (\bibinfo {year}
  {1993})}\BibitemShut {NoStop}%
\bibitem [{\citenamefont {Nazarov}\ and\ \citenamefont
  {Blanter}(2009)}]{QuantumTransport}%
  \BibitemOpen
  \bibfield  {author} {\bibinfo {author} {\bibfnamefont {Y.~V.}\ \bibnamefont
  {Nazarov}}\ and\ \bibinfo {author} {\bibfnamefont {Y.~M.}\ \bibnamefont
  {Blanter}},\ }\href@noop {} {\emph {\bibinfo {title} {Quantum transport:
  Introduction to Nanoscience}}}\ (\bibinfo  {publisher} {Cambridge University
  Press},\ \bibinfo {address} {Cambridge},\ \bibinfo {year} {2009})\BibitemShut
  {NoStop}%
\bibitem [{sup()}]{supplmat}%
  \BibitemOpen
  \href@noop {} {}\bibinfo {note} {See Supplemental Material at [URL will be
  inserted by publisher] for a detailed explanation of all activities of this
  project.}\BibitemShut {Stop}%
\bibitem [{\citenamefont {Wilkinson}\ and\ \citenamefont
  {Austin}(1993)}]{Wilkinson1}%
  \BibitemOpen
  \bibfield  {author} {\bibinfo {author} {\bibfnamefont {M.}~\bibnamefont
  {Wilkinson}}\ and\ \bibinfo {author} {\bibfnamefont {E.~J.}\ \bibnamefont
  {Austin}},\ }\bibfield  {title} {\bibinfo {title} {Densities of degeneracies
  and near-degeneracies},\ }\href {https://doi.org/10.1103/PhysRevA.47.2601}
  {\bibfield  {journal} {\bibinfo  {journal} {Phys. Rev. A}\ }\textbf {\bibinfo
  {volume} {47}},\ \bibinfo {pages} {2601} (\bibinfo {year}
  {1993})}\BibitemShut {NoStop}%
\bibitem [{\citenamefont {Walker}\ and\ \citenamefont
  {Wilkinson}(1995)}]{Wilkinson2}%
  \BibitemOpen
  \bibfield  {author} {\bibinfo {author} {\bibfnamefont {P.~N.}\ \bibnamefont
  {Walker}}\ and\ \bibinfo {author} {\bibfnamefont {M.}~\bibnamefont
  {Wilkinson}},\ }\bibfield  {title} {\bibinfo {title} {Universal fluctuations
  of chern integers},\ }\href {https://doi.org/10.1103/PhysRevLett.74.4055}
  {\bibfield  {journal} {\bibinfo  {journal} {Phys. Rev. Lett.}\ }\textbf
  {\bibinfo {volume} {74}},\ \bibinfo {pages} {4055} (\bibinfo {year}
  {1995})}\BibitemShut {NoStop}%
\bibitem [{\citenamefont {Walker}\ \emph {et~al.}(1996)\citenamefont {Walker},
  \citenamefont {S{\'a}nchez},\ and\ \citenamefont {Wilkinson}}]{Wilkinson3}%
  \BibitemOpen
  \bibfield  {author} {\bibinfo {author} {\bibfnamefont {P.~N.}\ \bibnamefont
  {Walker}}, \bibinfo {author} {\bibfnamefont {M.~J.}\ \bibnamefont
  {S{\'a}nchez}},\ and\ \bibinfo {author} {\bibfnamefont {M.}~\bibnamefont
  {Wilkinson}},\ }\bibfield  {title} {\bibinfo {title} {Singularities in the
  spectra of random matrices},\ }\href@noop {} {\ \textbf {\bibinfo {volume}
  {37}},\ \bibinfo {pages} {5019} (\bibinfo {year} {1996})}\BibitemShut
  {NoStop}%
\bibitem [{\citenamefont {Altland}\ and\ \citenamefont
  {Zirnbauer}(1997)}]{Altland}%
  \BibitemOpen
  \bibfield  {author} {\bibinfo {author} {\bibfnamefont {A.}~\bibnamefont
  {Altland}}\ and\ \bibinfo {author} {\bibfnamefont {M.~R.}\ \bibnamefont
  {Zirnbauer}},\ }\bibfield  {title} {\bibinfo {title} {Nonstandard symmetry
  classes in mesoscopic normal-superconducting hybrid structures},\ }\href
  {https://doi.org/10.1103/PhysRevB.55.1142} {\bibfield  {journal} {\bibinfo
  {journal} {Phys. Rev. B}\ }\textbf {\bibinfo {volume} {55}},\ \bibinfo
  {pages} {1142} (\bibinfo {year} {1997})}\BibitemShut {NoStop}%
\bibitem [{\citenamefont {Berry}\ and\ \citenamefont
  {Shukla}(2020)}]{Berry2020}%
  \BibitemOpen
  \bibfield  {author} {\bibinfo {author} {\bibfnamefont {M.~V.}\ \bibnamefont
  {Berry}}\ and\ \bibinfo {author} {\bibfnamefont {P.}~\bibnamefont {Shukla}},\
  }\bibfield  {title} {\bibinfo {title} {Geometric phase curvature
  statistics},\ }\href {https://doi.org/10.1007/s10955-019-02400-6} {\bibfield
  {journal} {\bibinfo  {journal} {Journal of Statistical Physics}\ }\textbf
  {\bibinfo {volume} {180}},\ \bibinfo {pages} {297} (\bibinfo {year}
  {2020})}\BibitemShut {NoStop}%
\bibitem [{\citenamefont {Gat}\ and\ \citenamefont
  {Wilkinson}(2021)}]{NewWilkinson}%
  \BibitemOpen
  \bibfield  {author} {\bibinfo {author} {\bibfnamefont {O.}~\bibnamefont
  {Gat}}\ and\ \bibinfo {author} {\bibfnamefont {M.}~\bibnamefont
  {Wilkinson}},\ }\bibfield  {title} {\bibinfo {title} {{Correlations of
  quantum curvature and variance of Chern numbers}},\ }\href
  {https://doi.org/10.21468/SciPostPhys.10.6.149} {\bibfield  {journal}
  {\bibinfo  {journal} {SciPost Phys.}\ }\textbf {\bibinfo {volume} {10}},\
  \bibinfo {pages} {149} (\bibinfo {year} {2021})}\BibitemShut {NoStop}%
\bibitem [{\citenamefont {Spivak}\ and\ \citenamefont {Zyuzin}(1991)}]{par1}%
  \BibitemOpen
  \bibfield  {author} {\bibinfo {author} {\bibfnamefont {B.}~\bibnamefont
  {Spivak}}\ and\ \bibinfo {author} {\bibfnamefont {A.}~\bibnamefont
  {Zyuzin}},\ }\bibfield  {title} {\bibinfo {title} {Chapter 2 - mesoscopic
  fluctuations of current density in disordered conductors},\ }in\ \href
  {https://doi.org/https://doi.org/10.1016/B978-0-444-88454-1.50008-5} {\emph
  {\bibinfo {booktitle} {Mesoscopic Phenomena in Solids}}},\ \bibinfo {series}
  {Modern Problems in Condensed Matter Sciences}, Vol.~\bibinfo {volume} {30},\
  \bibinfo {editor} {edited by\ \bibinfo {editor} {\bibfnamefont
  {B.}~\bibnamefont {Althshuler}}, \bibinfo {editor} {\bibfnamefont
  {P.}~\bibnamefont {Lee}},\ and\ \bibinfo {editor} {\bibfnamefont
  {R.}~\bibnamefont {Webb}}}\ (\bibinfo  {publisher} {Elsevier},\ \bibinfo
  {year} {1991})\ pp.\ \bibinfo {pages} {37--80}\BibitemShut {NoStop}%
\bibitem [{\citenamefont {Feng}(1991)}]{par2}%
  \BibitemOpen
  \bibfield  {author} {\bibinfo {author} {\bibfnamefont {S.}~\bibnamefont
  {Feng}},\ }\bibfield  {title} {\bibinfo {title} {Chapter 4 - conductance
  fluctuations and 1/f noise magnitudes in small disordered structures:
  Theory},\ }in\ \href
  {https://doi.org/https://doi.org/10.1016/B978-0-444-88454-1.50010-3} {\emph
  {\bibinfo {booktitle} {Mesoscopic Phenomena in Solids}}},\ \bibinfo {series}
  {Modern Problems in Condensed Matter Sciences}, Vol.~\bibinfo {volume} {30},\
  \bibinfo {editor} {edited by\ \bibinfo {editor} {\bibfnamefont
  {B.}~\bibnamefont {Altshuler}}, \bibinfo {editor} {\bibfnamefont
  {P.}~\bibnamefont {Lee}},\ and\ \bibinfo {editor} {\bibfnamefont
  {R.}~\bibnamefont {Webb}}}\ (\bibinfo  {publisher} {Elsevier},\ \bibinfo
  {year} {1991})\ pp.\ \bibinfo {pages} {107--129}\BibitemShut {NoStop}%
\bibitem [{\citenamefont {Campagnano}\ and\ \citenamefont
  {Nazarov}(2006)}]{Campagnano}%
  \BibitemOpen
  \bibfield  {author} {\bibinfo {author} {\bibfnamefont {G.}~\bibnamefont
  {Campagnano}}\ and\ \bibinfo {author} {\bibfnamefont {Y.~V.}\ \bibnamefont
  {Nazarov}},\ }\bibfield  {title} {\bibinfo {title} {${G}_{Q}$ corrections in
  the circuit theory of quantum transport},\ }\href
  {https://doi.org/10.1103/PhysRevB.74.125307} {\bibfield  {journal} {\bibinfo
  {journal} {Phys. Rev. B}\ }\textbf {\bibinfo {volume} {74}},\ \bibinfo
  {pages} {125307} (\bibinfo {year} {2006})}\BibitemShut {NoStop}%
\bibitem [{\citenamefont {Beenakker}(1997)}]{Beenakker}%
  \BibitemOpen
  \bibfield  {author} {\bibinfo {author} {\bibfnamefont {C.~W.~J.}\
  \bibnamefont {Beenakker}},\ }\bibfield  {title} {\bibinfo {title}
  {Random-matrix theory of quantum transport},\ }\href
  {https://doi.org/10.1103/RevModPhys.69.731} {\bibfield  {journal} {\bibinfo
  {journal} {Rev. Mod. Phys.}\ }\textbf {\bibinfo {volume} {69}},\ \bibinfo
  {pages} {731} (\bibinfo {year} {1997})}\BibitemShut {NoStop}%
\bibitem [{\citenamefont {Repin}\ \emph {et~al.}(2019)\citenamefont {Repin},
  \citenamefont {Chen},\ and\ \citenamefont {Nazarov}}]{Repin}%
  \BibitemOpen
  \bibfield  {author} {\bibinfo {author} {\bibfnamefont {E.}~\bibnamefont
  {Repin}, \bibfnamefont {V}}, \bibinfo {author} {\bibfnamefont
  {Y.}~\bibnamefont {Chen}},\ and\ \bibinfo {author} {\bibfnamefont
  {Y.}~\bibnamefont {Nazarov}, \bibfnamefont {V}},\ }\bibfield  {title}
  {\bibinfo {title} {{Topological properties of multiterminal superconducting
  nanostructures: Effect of a continuous spectrum}},\ }\bibfield  {journal}
  {\bibinfo  {journal} {{Phys. Rev. B}}\ }\textbf {\bibinfo {volume} {{99}}},\
  \href {https://doi.org/{10.1103/PhysRevB.99.165414}}
  {{10.1103/PhysRevB.99.165414}} (\bibinfo {year} {{2019}})\BibitemShut
  {NoStop}%
\bibitem [{\citenamefont {van Heck}\ \emph {et~al.}(2014)\citenamefont {van
  Heck}, \citenamefont {Mi},\ and\ \citenamefont {Akhmerov}}]{Akhmerov}%
  \BibitemOpen
  \bibfield  {author} {\bibinfo {author} {\bibfnamefont {B.}~\bibnamefont {van
  Heck}}, \bibinfo {author} {\bibfnamefont {S.}~\bibnamefont {Mi}},\ and\
  \bibinfo {author} {\bibfnamefont {A.~R.}\ \bibnamefont {Akhmerov}},\
  }\bibfield  {title} {\bibinfo {title} {Single fermion manipulation via
  superconducting phase differences in multiterminal josephson junctions},\
  }\href {https://doi.org/10.1103/PhysRevB.90.155450} {\bibfield  {journal}
  {\bibinfo  {journal} {Phys. Rev. B}\ }\textbf {\bibinfo {volume} {90}},\
  \bibinfo {pages} {155450} (\bibinfo {year} {2014})}\BibitemShut {NoStop}%
\bibitem [{\citenamefont {Chen}\ and\ \citenamefont {Nazarov}(2021)}]{Chen}%
  \BibitemOpen
  \bibfield  {author} {\bibinfo {author} {\bibfnamefont {Y.}~\bibnamefont
  {Chen}}\ and\ \bibinfo {author} {\bibfnamefont {Y.~V.}\ \bibnamefont
  {Nazarov}},\ }\bibfield  {title} {\bibinfo {title} {Spintronics with a weyl
  point in superconducting nanostructures},\ }\href
  {https://doi.org/10.1103/PhysRevB.103.165424} {\bibfield  {journal} {\bibinfo
   {journal} {Phys. Rev. B}\ }\textbf {\bibinfo {volume} {103}},\ \bibinfo
  {pages} {165424} (\bibinfo {year} {2021})}\BibitemShut {NoStop}%
\bibitem [{\citenamefont {Barakov}\ and\ \citenamefont
  {Nazarov}(2021)}]{Hristo2021Dec}%
  \BibitemOpen
  \bibfield  {author} {\bibinfo {author} {\bibfnamefont {H.}~\bibnamefont
  {Barakov}}\ and\ \bibinfo {author} {\bibfnamefont {Y.}~\bibnamefont
  {Nazarov}},\ }\bibfield  {title} {\bibinfo {title} {{Abundance of Weyl points
  in semiclassical multi-terminal superconducting nanostructures [Data set]}},\
  }\bibfield  {journal} {\bibinfo  {journal} {Zenodo}\ }\href
  {https://doi.org/10.5281/zenodo.5806468} {10.5281/zenodo.5806468} (\bibinfo
  {year} {2021})\BibitemShut {NoStop}%
\end{thebibliography}%

\newpage
\renewcommand{\theequation}{S\arabic{equation}}
\setcounter{equation}{0}
\renewcommand{\thefigure}{S\arabic{figure}}
\renewcommand{\figurename}{Supplementary Fig.}

\setcounter{figure}{0}
\renewcommand{\thesection}{S\arabic{section}}
\setcounter{section}{0}

\begin{widetext}
\section*{Supplemental Material}
In this Supplemental Material, we present additional details about the activites A, B, and C. 

\section{Activity A: density and charge correlations of WP in a uniform parameter space\label{Sec:ActA}}
The goal of this activity is to relate the actual density of WP's with the universal correlation parameter $l_c$. We implement a variation of Wilkinson model with random class C $2N \times 2N$ matrices (Eq. 1 of the main text) where $l_c \equiv \pi \sqrt{3/2N}$ does not depend on the position in the 3d space of the phases.
N=60 lc = 0.496  N=80 lc = 0.430 

To find the WP's, we implement an iterative optimization procedure in the 3D parameter space. The optimization function is the smallest in modulus eigenvalue of the matrix.
Since it involves the matrix diagonalization, the computation time scales as $N^3$. The initial position is chosen randomly. The coordinates of a WP are found after several tens of iterations. To find {\it all} WP's, we repeat the procedure again and again, keeping the list of WP's found to exclude the duplicates. We learned from the experience that the procedure has to be repeated five times the expected number of WP's: further runs do not deliver new points. Since the expected number of points scales as $N^{3/2}$, the total computation time scales as $N^{9/2}$ and really big $N$ are not accessible for practical calculations. The following table summarizes our concrete results for the number of $WP$ averaged for a number of runs :
\begin{center}
	\begin{tabular}{||c c c c||} 
		\hline
		N &  $40$ &  $60$ &  $80$ \\
		\hline\hline
		$\#$ runs & $34$ & $5$ & $2$ \\ 
		\hline
		$<N_{W}>$ & $930.5 \pm 60.5$ & $1692.0 \pm 49.9 $ & $2571.5 \pm 24.5$ \\
		\hline
		 $c_{W}$ & $0.844 \pm 0.054$ & $0.836 \pm 0.024$ & $0.825 \pm 0.007$  \\
		\hline
	\end{tabular}
\end{center}
This brought us to the value $c_{W}=0.83 \pm 0.05$ (Eq. 2 of the main text).

As an extra check of the method in use, we compute the concentration of the level crossings far from zero energy, namely between the 10th and 11th level. We expect this to be close to the concentration of WP's in the GUE ensemble. Willkinson et al. have computed this concentration to be $(2/3)\sqrt{\pi} \approx 1.18 $ in units of $l_c^{-3}$. Our calculation for $N=40$ averaged over 6 runs gives a consistent value $1.146 \pm 0.065$. 

In the course of calculations, we have accumulated significant statistics of WP coordinates and their charges. We hoped that these statistics suffices to compute the charge-charge correlations of WP distribution. However, this did not work.
The histograms approximating the charge density at a given distance from a WP exhibited significant fluctuations at relatively large distances. Our attempts to smooth these fluctuations considering the Laplace transform of the charge-charge correlator initially led us to an erroneous conclusion of a power-like tail in this correlator. 

Fortunately, we checked these conclusions with an alternative method. We have computed the correlator of Berry curvatures at given distances. The most general form of the correlator of two vector quantities $B^\alpha$ in the dimensionless coordinates $\vec{r}$ reads 
\begin{equation}
\langle\langle B^{\alpha}(\vec{R}) B^\beta(\vec{R}+\vec{r}) \rangle\rangle = \delta_{\alpha\beta} B(r) + \frac{r^\alpha r^\beta}{r^2} B_1(r)
\end{equation}
Since $Q(\vec{r}) = (4\pi)^{-1} \partial_\alpha B^\alpha(\vec{r})$, the charge-charge correlator is then expressed as
\begin{equation}
(4\pi)^2 \langle\langle Q(\vec{R}) Q(\vec{R}+\vec{r}) \rangle\rangle = \frac{3}{r^2} \frac{\partial}{\partial r}\left(r B_1(r) \right) - \frac{1}{r^2}\frac{\partial}{\partial r}\left(r^2 \frac{\partial}{\partial r} B(r)\right)
\end{equation}

We compute the correlator at each $r$ separately accumulating the statistics of Berry curvatures $B_{1,2}^\alpha$ in two random points $1,2$ separated by the distance $r$. The computation is relatively fast so for each point we can accumulate $10^4$ samples for $N=40$ and $10^5$ samples for $N=20$. It may seem that the two independent functions in the correlator are just given by the average products 
\begin{align}
B(r) &=& \frac{1}{4}\left(\langle B^\alpha_1  B^\alpha_2 \rangle - \langle B^\alpha_1 n^\alpha B^\beta_2 n^\beta\rangle\right); \\
B_1(r) &=& -\frac{1}{2}\langle B^\alpha_1  B^\alpha_2 \rangle + \frac{3}{2}  \langle B^\alpha_1 n^\alpha B^\beta_2 n^\beta\rangle.
\end{align}
However, the evaluation is not so simple. As has been mentioned in \cite{Berry2020}, the distribution of $B^\alpha$ has long power-law tails resulting in an infinite variance. Owing to this, the accuracy of computed averages does not increase with the number of samples in the statistics. 

The universal prescription to evaluate the averages in this situation is to disregard the large values. We implement it in the following fashion: we rescale the accumulated values of $B^\alpha$ to decrease it if large,
\begin{equation}
\bar{B}^\alpha = \frac{B^{\alpha}}{\sqrt{1+B^{\beta} B^{\beta}/B^2_0}}.
\end{equation}
Here, $B_0$ is a parameter and the modulus of the rescaled $\bar{B}^\alpha$ never exceeds $B_0$. The variance of $\bar{B}^\alpha$ is thus finite, and usual statistical considerations do apply. The value of $B_0$ should not be taken too small since the averages would not approximate the correlator, nor too large since the large values of $B^\alpha$ would not be suppressed. In practice, we plot the averages versus $B_0$ and pick up the value of the average that persist in a large interval of $B_0$.

Within statistical error, $B_1(r)=0$ for all r. We can prove this analytically for $r \ll 1$. With this, 
\begin{equation}
\langle\langle Q(0) Q(\vec{r})\rangle\rangle = (4\pi)^{-2} \nabla^2 B(r).
\end{equation}
AS stated in the main text, the correlator of Berry curvatures can be approximated with
\begin{equation}
B(r)\approx 10.4 e^{-2.8 r - 3.3 r^2}  
\end{equation}
This has no trace of long-distance power-law correlations.

This expression also proves the electro-neutrality of the WP distribution,
\begin{equation}
\int d\vec{r} \langle\langle Q(0) Q(\vec{r})\rangle\rangle = -N_w/V
\end{equation}
that is, the presence of a WP with the charge $+1$ at a point results in a depletion of average charge density around the point, the charge depleted being $-1$.  

After completion of these calculation, we became aware of a similar calculation of the Berry curvature correlations. \cite{NewWilkinson}. The authors address the correlator in a general GUE ensemble that is similar but distinct from near-zero energy correlator of interest. However, they use the same fitting function and end up with similar coefficients. In our notations, they give
\begin{equation}
r B(r) \approx 7.42 e^{- 3.56 x - 2.03 x^2}.
\end{equation}

\section{Activity B: finding $l_c$ in quantum circuit theory \label{Sec:ActB}}

The goal of this activity is to find the scale governing universal parametric correlations for concrete nanostructures that can modelled with quantum circuit theory \cite{QuantumTransport}. For a 3D parameter space, this scale is defined as   
\begin{equation}
l_c^{-3} = \sqrt{{\rm det}\langle\langle v_i v_j \rangle \rangle}/\delta^3_S \end{equation}
where  $v_i \equiv \partial E /\partial \phi_i$ is the "velocity" of an energy level in the spectrum and  $\delta_S$ is the mean level spacing. The parameter $l_c^{-3}$ depends on energy as well as on a point in the parameter space. 

\subsection{ The action, mean level spacing, and the velocity correlator} At quantum level, the nanostructure is characterized by an electron scattering matrix $S$ in the space of of the quantum channels incoming from the leads where the superconducting phase is incorporated with a factor $e^{i\phi_i}$ ascribed to a channel coming from the lead $i$. One can derive (see e.g. \cite{Repin}) an action for imaginary-time Green functions characterizing the nanostructure,
\begin{equation}
{\cal S}(\epsilon) = -\frac{1}{2} {\rm Tr} {\rm ln} \left(\frac{E+\epsilon}{2E} + \frac{E-\epsilon}{2E} S S^*\right)
\end{equation}
$\epsilon$ being the imaginary energy, $E\equiv \sqrt{\epsilon^2 +\Delta^2}$. It is known \cite{Beenakker} that the Andreev bound states are related to the eigenvalues of $S S^*$. For these eigenvalues, we will use
\begin{equation}
S S^* \rightarrow - e^{i \lambda}.
\end{equation}
The eigenvalues $\lambda$ come in $\pm$ pairs.
The energies of the bound states correspond to zeros of the expression under the log. We introduce the notation
$\epsilon/E = \sin \theta$, $-\pi/2<\theta<\pi/2$, and  rewrite the action as
\begin{equation}
{\cal S}(\theta) = -\sum_{\lambda} {\rm ln} \left( 1-\cos^2\theta\cos^2(\lambda/2)\right)
\end{equation}

Since we are interested in characteristics of the spectrum near zero energy, we can expand in small $\theta, \lambda$ so the action becomes:
\begin{equation}
{\cal S} = -\sum_\lambda {\rm ln} \left( \theta - i \lambda/2 \right) = -\sum_{\lambda>0} {\rm ln} \left( \theta^2 +\lambda^2/4 \right)
\end{equation}
In this limit, each $\lambda$ gives a bound state at energy 
$\Delta \lambda$.
Let us compute the derivative of the action with respect to $\theta$:
\begin{equation}
\frac{\partial {\cal S}}{\partial \theta} = - \sum_{\lambda>0} \frac{2\theta}{\theta^2 +\lambda^2/4} = - \rho_{\lambda} \int_{0}^{\infty} d{\lambda}\frac{2\theta}{\theta^2 +\lambda^2/4} = -2\pi {\rm sgn} (\theta) \rho_{\lambda} 
\label{eq:forrho}
\end{equation}
where we made a semiclassical approximation replacing the summation over the discrete $\lambda$ with integration over their continuous density $\rho_\lambda$. The semiclassical action has therefore a cusp at $\theta=0$, the value of the cusp determines the density of the eigenvalues, that is directly related to the mean level spacing.

Let us look at the random velocities of the levels, the velocity with respect to a parameter $\alpha$ being $v_{\alpha} \equiv \partial_\alpha \lambda$
The derivative of the action then reads
\begin{equation}
\partial_\alpha {\cal S} = - \sum_{\lambda>0} \frac{\lambda v_{\alpha}}{2(\theta^2 +\lambda^2/4)}
\end{equation}
The velocities correlate at the same level only. For the correlator of the derivatives, this gives 
\begin{equation}
\label{eq:correlator}
\langle\langle \partial_\alpha {\cal S}(\theta) \partial_\beta {\cal S}(\theta') \rangle \rangle  = \langle\langle v_\alpha v_\beta  \rangle \rangle \frac{\rho_\lambda}{4} \int_0^{\infty}\frac{d\lambda \ \lambda^2}{(\theta^2 +\lambda^2/4) (\theta'^2 +\lambda^2/4)} = \pi \frac{\langle\langle v_\alpha v_\beta  \rangle \rangle \rho_\lambda}{|\theta|+|\theta'|}
\end{equation}
This implies that if we know $\langle\langle  {\cal S}(\theta,\vec{\phi}) {\cal S}(\theta',\vec{\phi}) \rangle \rangle$, we can evaluate the velocity correlators and $l_c^{-3}$

\subsection{Semiclassics: saddle point}
In quantum circuit theory approach, the same action is expressed as a functional of the matrices $\check{G}$, $\check{G}^2=1$, ${\rm Tr} \check{G}=0$ that is defined in the nodes and reservoirs of the nanostructure. The situation in hand we can describe with $2\times 2$ matrices. These matrices are fixed in the leads
\begin{equation}
\label{eq:reservoir}
\check{G}_i = \begin{bmatrix}
\sin \theta & \cos \theta e^{-i\phi_i}\\
\cos \theta e^{i\phi_i}   & -\sin \theta
\end{bmatrix}.
\end{equation} 
The matrices in the nodes are obtained by minimization of the action. In case of a short nanosctructure, the action is a sum of contributions of each connector,
\begin{equation}
{\cal S} = \sum_{c} {\cal S}_c; \; {\cal S}_c = \frac{1}{2} {\rm Tr} {\cal F}_c \left(\frac{1}{2}(\check{G}_a \check{G}_b + \check{G}_b \check{G}_a)\right).
\end{equation}
The function ${\cal F}_c$ is proportional to the conductance of the connector and depends on the type of the connector. For instance, for a quantum point contact of conductance $G$ ${\cal F}(x) = -(G/G_Q) \ln((x+1)/2)$. Generally, a connector is characterized by a set of transmission coefficients $T_p$, and 
\begin{equation}
{\cal F}(x) = - \sum_p \ln\left(1 + \frac{T}{2}(x-1) \right).
\end{equation}
Long nanostructures can be 
described with addition of "leakage" connectors \cite{QuantumTransport}, yet we do not need this extension since the density of Weyl points is determined by the spectrum properties at zero energy that do not depend on the size of the nanostructure.

We will restrict ourselves to the simplest situation with a single node in the nanostructure and any number of the reservoirs. The connectors can be labelled with the lead index $i$, and
\begin{equation}
{\cal S} = \frac{1}{2} \sum_i {\rm Tr} {\cal F}_i \left(\frac{1}{2}(\check{G}_i \check{G} + \check{G} G_i)\right)
\end{equation}
where $\check{G}$ is the matrix in the node. We will make use of the fact that $\check{G}_i \check{G} + \check{G} \check{G}_i$ is a number rather than a matrix for any $2\times 2$ matrices.

It is constructive to map the $2\times 2$ matrices on the corresponding 3D vectors,
\begin{equation}
\check{G}_i \rightarrow (\cos\theta \cos\phi_i, \cos\theta \sin\phi_i,\sin\theta )
\end{equation}
while the node matrix
\begin{equation}
\check{G} \rightarrow (\cos\Theta \cos\Phi, \cos\Theta \sin\Phi, \sin \Theta)
\end{equation}
This gives the following inner products for each connector,
\begin{equation}
s_i = \sin\theta \sin \Theta + \cos\Theta \cos\theta \cos(\phi_i - \Phi),
\end{equation}
so the action reads
\begin{equation}
{\cal S} = \sum_i {\cal F}_i(s_i) 
\end{equation}
To find $\check{G}$, we minimize with respect to $\Theta$, $\phi$. This gives two conditions:
\begin{eqnarray}
0&=&\sum_i {\cal F}'_i(s_i) \partial_{\Theta} s_i = \sum_i  {\cal F}'_i(s_i) (\sin\theta \cos\Theta -\sin\Theta \cos(\phi_i - \Phi))\\
0&=&\sum_i {\cal F}'_i(s_i) \partial_{\Phi} s_i = \sum_i  {\cal F}'_i(s_i) \sin(\phi_i - \Phi)
\end{eqnarray}
In the limit of $\theta \to 0$, this becomes
\begin{eqnarray}
0&=&\sum_i  {\cal F}'_i(s_i) \cos(\phi_i - \Phi) = \sum_i  {\cal F}'_i(s_i) s_i\\
0&=&\sum_i {\cal F}'_i(s_i) \sin(\phi_i - \Phi)
\end{eqnarray}
We can extract level spacing from the cusp of the action at small $\theta$. By virtue of optimization, ${\rm sgn} \Theta = {\rm sgn} \theta$. Therefore, at small values of $\theta$ $s_i \rightarrow s_i + |\theta| |\sin \Theta|$, and the cusp part of the action reads
\begin{equation}
{\cal S} = |\theta||\sin \Theta| \sum_i {\cal F}'_i(s_i) 
\end{equation}
The density of the eigenvalues is then extracted with the aid of Eq. \ref{eq:forrho}.

\subsection{Semiclassics: correlations}
To compute the correlations of the action at two different parameter settings, $(\theta, \phi_i)$, and $(\theta', \phi'_i)$,  we have to double the dimension of the matrices. So we consider $4\times 4$ matrices. We need to do this separately for diffusion and Cooperon channels \cite{Campagnano}. For the reservoirs, these matrices are made from two diagonal blocks, each corresponding to a setting of the corresponding reservoir. We will distinguish the settings marking or not marking then with a prime,
\begin{equation}
\check{G}_i \rightarrow \begin{bmatrix}
G_i & 0\\
0   & G'_i
\end{bmatrix}
\end{equation}
For the diffusion channel, $G'_i$ is just given by Eq. \ref{eq:reservoir}. For the Cooperon channel, $G'_i$ is {\it transposed}.

The optimization of the action results in the block-diagonal matrix in the node
\begin{equation}
\hat{G}_0 \rightarrow \begin{bmatrix}
G_0 & 0\\
0   & G'_0
\end{bmatrix}
\end{equation}
To compute the correlations, we have to derive the quadratic expansion of the action near this optimum. With the quadratic accuracy, $\hat G$ is given by 
\begin{equation}
\check{G} = \check{G}_0 + \check{g} - \frac{1}{2} \check{G}_0 \check{g}^2; \; \check{g}\check{G} + \check{G}\check{g} =0
\end{equation}
We need to substitute this to the action and expand it to the terms quadratic in $\check{g}$. The first-order terms cancel since $\check{G}_0$ corresponds to the minimum of the action.
 
This calculation is made most efficiently in the basis where $G_0$ is a diagonal matrix, 
\begin{equation}
\hat{G}_0 = \begin{bmatrix}
1&0& 0& 0 \\
0&-1&0& 0 \\
0& 0&1& 0 \\
0& 0&0& -1
\end{bmatrix}
\end{equation}
In this basis, a matrix of a reservoir read
\begin{equation}
\hat{G}_i = \begin{bmatrix}
s&u& 0& 0 \\
u^*&-s&0& 0 \\
0& 0&s'& u' \\
0& 0&u'^*& -s'
\end{bmatrix}\end{equation}
where $s,u$ posess the index $i$, $s^2+ |u|^2=1$, $s$ being inner product introduced earlier. The minimization equation in these basis reads:
\begin{equation}
0= \sum_i {\cal F}'(s_i) u_i. 
\end{equation}

Let us specify those more explicitly. We choose a basis in the space of 3d vectors with the z-axis in the direction of $G_0$, the angle $\phi$ is counted from $\Phi$,
\begin{eqnarray}
\vec{x} = (-\sin\Theta, 0, \cos\Theta)\\
\vec{y} = (0,1,0)\\
\vec{z} = (\cos\Theta, 0, \sin(\Theta))
\end{eqnarray}
From this, 
\begin{equation}
u = -\sin\Theta\cos\theta\cos(\phi_i-\Phi) + \cos\Theta\sin\theta + i \cos(\theta)\sin(\phi_i -\Phi).
\end{equation}
At $\theta \to 0$, 
\begin{equation}
u =  - \sin\Theta\cos(\phi_i-\Phi) + i \sin(\phi_i -\Phi).
\end{equation}

Let us proceed with the expansion. We need to choose the $\check{g}$ in the non-diagonal blocks and guarantee that it anti-commutes with $\check{G}$ - otherwise, it will not modify $\check{G}$ and the value of the action will not change. The most general matrix of this kind can be parametrized as
\begin{equation}
\begin{bmatrix}
0&0& 0& w_2 \\
0&0&w_1& 0 \\
0& v_1&0&0 \\
v_2& 0&0&0
\end{bmatrix}
\end{equation}
A straightforward but lengthy calculation results in the following form: 
\begin{equation}
\delta{\cal S} = \frac{1}{2}\sum_i \begin{bmatrix}v_1\\v_2\end{bmatrix}
\begin{bmatrix}
A_i & B^*_i \\
B_i & A_i 
\end{bmatrix}
\begin{bmatrix}w_1\\w_2\end{bmatrix}
\end{equation}
with 
\begin{equation}
A_i =  \frac{{\cal F}'_i(s_i) (1-s^2_i) - {\cal F}'_i(s'_i) (1-s'^2_i)}{s_i-s'_i};\; 
B_i = u_i u'_i\frac{{\cal F}'(s_i)-{\cal F}'(s'_i)}{s_i-s'_i}
\end{equation}
This expression is for the diffusion channel, for the Cooperon channel we need to replace $u' \rightarrow u'^*$.

\subsection{Close points}
For our task, we need to analyse the quadratic form in close points, $\phi_i \to \phi'_i$. The form of the correlator given by Eq. \ref{eq:correlator} suggests that there is an eivenvalue of this matrix that goes to $0$ at $\phi_i \to \phi'_i$ and $\theta \to 0$, and the parametric dependence of this eigenvalue defines the correlations. The calculation shows  no such eigenvalue in the diffusion channel, so from now on we concentrate on the Cooperon one.
  In this case, 
\begin{equation}
A = \sum_i ({\cal F}'(s_i) (1-s^2_i))' = \sum_i {\cal F}''(s_i) (1-s_i^2) -2 \sum_i {\cal F}'(s_i) s_i; 
B = \sum_i {\cal F}''(s_i) (1-s_i^2)
\end{equation}
At $\theta \to 0$, $\sum_i {\cal F}'(s_i) s_i \to 0$ and the determinant $A^2 - |B^2|$ vanishes indicating the small eigenvalue expected.
The determinant can be presented as
\begin{equation}
(A - {\rm Re} B)(A +{\rm Re} B) - ({\rm Im} B)^2
\end{equation}
where $A - {\rm Re} B$ goes to zero in close  points at $\theta \to 0$, $A + {\rm Re} B$ does not, 
and ${\rm Im} B$ goes to zero in close points irrespective of $\theta$.

Let us compute $A-B$ in conciding points.
\begin{equation}
A = ({\cal F}'(s)(1-s^2))'; \; B = (1-s^2) {\cal F}''(s) \rightarrow A-B = - 2 s{\cal F}'(s)
\end{equation}
Here, the summation over $i$ is implied.  We know that $\sum_i {\cal F}'(s_i) u_i =0$ at any $\theta$. 
We note that 
\begin{eqnarray}
s_i &=& \sin \theta \sin \Theta + \cos\theta\cos\Theta \cos(\phi_i -\Phi)\\
{\rm Re}(u_i) &=& - \sin\Theta \cos\theta \cos(\phi_i -\Phi) +\sin\theta \cos\Theta
\end{eqnarray}
To this end, we evaluate
\begin{equation}
\sum_i {\cal F}'(s_i) s_i = \sum_i {\cal F}'(s_i) (s_i + {\rm cotan} \Theta {\rm Re} u_i) = \frac{\sin\theta}{\sin\Theta}\sum_i {\cal F}'(s_i) \end{equation}
Generaling to small $\theta$ and the same points in the phase space, we obtain
\begin{equation}
A-B = - \frac{|\theta| +|\theta'|}{\sin\Theta} \sum_i {\cal F}'(s_i)
\end{equation}
Let us compute $A+B$. We can neglect $\theta$ and the difference between the points to obtain
\begin{equation}
A+B = 2 \sum_i {\cal F}''(s_i) (1-s_i^2)
\end{equation}
Let us compute the terms in $A - {\rm Re} B$ that are proportional to the squares of the differences between the points.
We start with
\begin{equation}
A - {\rm Re} B = \frac{{\cal F}'(s) (1-s^2) - {\cal F}'(s') (1-s'^2)}{s-s'} - \frac{ u u'^* + u'u^*}{2} \frac{{\cal F}'(s)-{\cal F}'(s')}{s-s'}
\end{equation}
Let us represent
\begin{equation}
u =\sqrt{1-s^2} e^{i\mu};\; u' =\sqrt{1-s'^2} e^{i\mu'}
\end{equation}
With this, the difference becomes
\begin{equation}
A - {\rm Re} B = \frac{{\cal F}'(s) (1-s^2) - {\cal F}'(s') (1-s'^2)}{s-s'} - \sqrt{1-s^2}\sqrt{1-s'^2} \cos(\mu-\mu') \frac{{\cal F}'(s)-{\cal F}'(s')}{s-s'}
\end{equation}
There are two contributions to the difference. One comes from $\delta \mu_i$ and reads
\begin{equation}
A - {\rm Re} B = \frac{1}{2} \sum_i (\delta \mu)^2_i (1-s_i^2) {\cal F}''(s_i)
\end{equation}
For another one, one can set $\delta \mu =0$.
\begin{eqnarray}
A - {\rm Re} B = \\
\frac{{\cal F}'(1-s^2) - {\cal F}'(s')(1-s'^2) + \sqrt{(1-s^2)(1-s'^2)} ({\cal F}'(s') - {\cal F}'(s))}{s-s'} =\\
- \frac{s+s'}{\sqrt{1-s^2} +\sqrt{1-s'^2}} \left({\cal F}'(s) \sqrt{1-s^2} + {\cal F}'(s') \sqrt{1-s^2} \right)
\end{eqnarray}
To simplify, we may add ${\cal F}'(s)s +{\cal F}'(s')s'$ that is zero under sum provided $\theta=0$.
This gives
\begin{equation}
A - {\rm Re} B =  \frac{\sqrt{(1-s^2)(1-s'2)}}{\sqrt{1-s^2} +\sqrt{1-s'^2}} ({\cal F}'(s) - {\cal F}'(s')) \left(\frac{s}{\sqrt{1-s^2}} - \frac{s'}{\sqrt{1-s'^2}}\right)
\end{equation}
With this, we get for the difference
\begin{equation}
A - {\rm Re} B  = \frac{1}{2} \sum_i (\delta s)_i \frac{{\cal F}''(s_i)}{1-s^2_i} + \frac{1}{2}\sum_i (\delta \mu)^2_i (1-s_i^2) {\cal F}''(s_i)
\end{equation}
We also have to inspect ${\rm Im} B$.
\begin{equation}
Im B = i \frac{uu'^*-u^*u}{2} \frac{{\cal F}'(s) - {\cal F}'(s')}{s-s'} \approx {\delta \mu} (1-s^2) {\cal F}''(s)
\end{equation}
With all this, 
\begin{eqnarray}
{\rm ln \ det} = {\rm ln} (C +D)\\
C = - \frac{|\theta| +|\theta'|}{\sin\Theta} \sum_i {\cal F}'(s_i) \\
D = \frac{1}{2} \sum_i (\delta s)_i \frac{{\cal F}''(s_i)}{1-s^2_i} + \frac{1}{2}\sum_i (\delta \mu)^2_i (1-s_i^2) {\cal F}''(s_i)\\ - \frac{\left(\sum_i{\delta \mu_i} (1-s_i^2) {\cal F}''(s_i)\right)^2}{2 \sum_i {\cal F}''(s_i) (1-s_i^2)}
\end{eqnarray}
The latter part can be presented as 
\begin{equation}
D = \frac{1}{2} D_{\alpha\beta} \delta \phi_\alpha \delta \phi_\beta
\end{equation}
$\alpha,\beta$ labelling the independent phases.
For this, we need to express $\delta s_i$, $\delta \mu_i$ in terms of $\delta \phi_i$. The corresponding formulas are straightforward but rather cumbersome. In fact, we do not use those in numerical calculations, but rather compute $\delta s_i$, $\delta \mu_i$ in terms of $\delta \phi_i$ to evaluate the quadratic form $D_{\alpha\beta}$. So we do not give these formulas here.

\subsection{Resulting relation}
We recall that the density in the phase space can be expressed as
\begin{equation}
l_c^{-3} =\sqrt{{\rm det}<v_\alpha v_\beta>} \rho^3_\lambda
\end{equation}
We have derived that 
\begin{eqnarray}
\rho_\lambda = \frac{|\sin \Theta|}{2\pi} |\sum_i {\cal F}'(s_i)| \\
{\cal S}_{\alpha\beta} = \pi \frac{\langle\langle v_\alpha v_\beta  \rangle \rangle \rho_\lambda}{|\theta|+|\theta'|} \\
{\cal S}_{\alpha\beta} = \frac{D^{\alpha\beta}}{C} 
\end{eqnarray}
So we get
\begin{equation}
\pi^2 <v_\alpha v_\beta> \rho^2_\lambda = D_{\alpha\beta} \frac{\sin^2\Theta}{2}
\end{equation}
Finally, collecting all terms, we obtain: 
\begin{equation}
l_c^{-3} = \sqrt{{\rm det} D_{\alpha\beta}} \frac{\sin^3\Theta}{2\sqrt{2}\pi^3}.
\end{equation}
We will use this formula for numerical evaluations. We stress this requires minimization for each set of $\phi_i$ to compute $\Theta, \Phi$ and minimization around this point to evaluate $D_{\alpha\beta}$.

\section{Activity C: WP positions for ballistic cross \label{Sec:ActC}}
Within the activity, we find the coordinates of the WP's in a ballistic cavity model connected to four superconducting leads. We take a random realization of the $4N \times 4N$ electron scattering matrix, augment it with the phases of the superconducting reservoirs and find the points where $SS^*$ has an eigenvalue $-1$ with the optimization procedure similar to that described in Section \ref{Sec:ActB}. 
Our results for the total number of points are summarized in the following table 
\begin{center}
	\begin{tabular}{||c c c c||} 
		\hline
		N &  $50$ &  $35$ &  $20$ \\
		\hline\hline
		$\#$ runs & $15$ & $8$ & $10$ \\ 
		\hline
		$<N_{W}>$ & $153.5\pm 13.8$ & $93.0\pm 11.8 $ & $46.4\pm 6.7$ \\
		\hline
	\end{tabular}
\end{center}
From this, we inherit $N_w = 0.40 G/G_Q$ as cited in the main text.
  
The points are found within the gapless region predicted by the semiclassical calculation. In Fig. \ref{fig:Supplemental-boundary} we present the boundaries of the gapless region for several cross-sections of the Brillouin zone in $\chi_{1,2,3}$ coordinates. We see that the boundary touches the centres of the squares and hexagons bounding the Brillouin zone. 

To check the correspondence of the positions of the WP's found with the predictions of the semi-classical theory, we compute the semi-classical density $l_c^{-3}$ in the positions found and accumulate the data to a histogram. The resulting distribution should differ  by a factor of $l_c^{-3}$ from the distribution of $l_c^{-3}$ itself. Indeed, when we plot together the distribution of $l_c^{-3}$ and the distribution corrected by the factor, we observe a satisfactory correspondence (Fig. 3c of the main text).

In conclusion, we present several 3D views of a realization of WP for $N=50$. (Fig. \ref{fig:Supplemental-points})

\begin{figure}
\centering
\includegraphics[width=\textwidth]{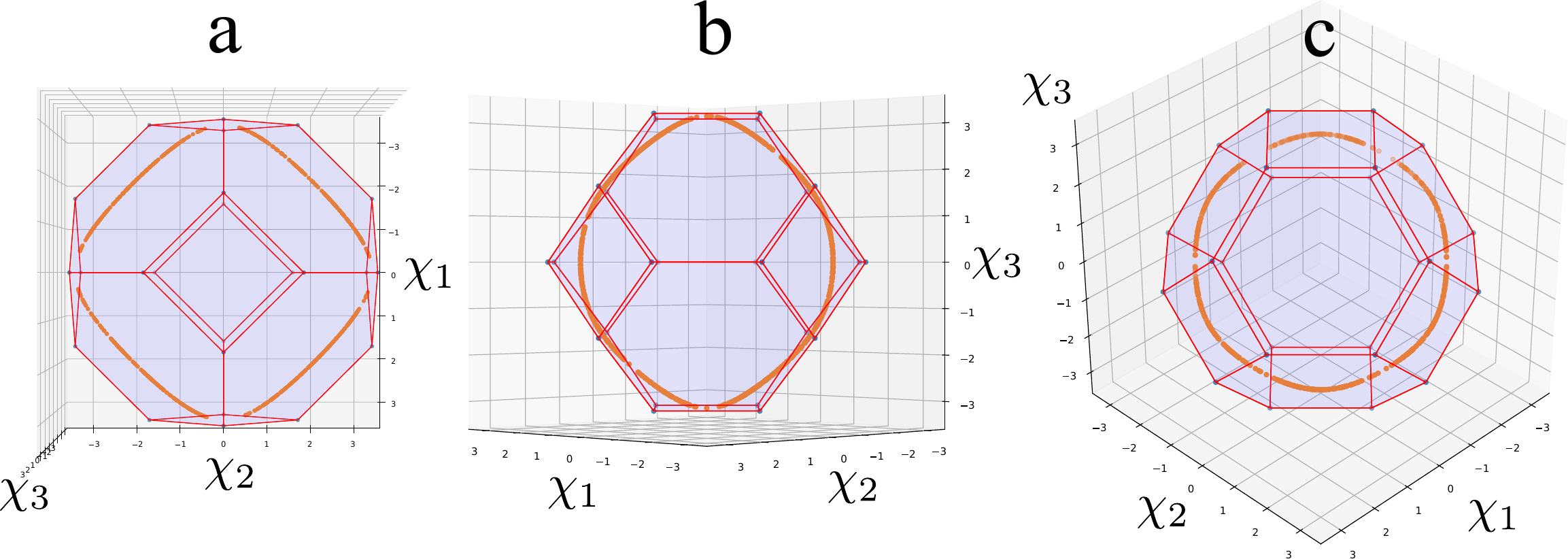}
\caption{The boundary of the gapless region (orange points) for several cross-sections plotted in a 3D along with the edges of the Brillouin zone. The view axis is perpendicular to the cross-section plane. a: cross-section plane $\chi_3=0$, b: cross-section plane $\chi_1=\chi_2$, c: cross-section plane $\chi_1+\chi_2+\chi_3=0$. }
\label{fig:Supplemental-boundary}
\end{figure}

\begin{figure}
	\begin{subfigure}{0.45\textwidth}
		\includegraphics[width=\textwidth]
		{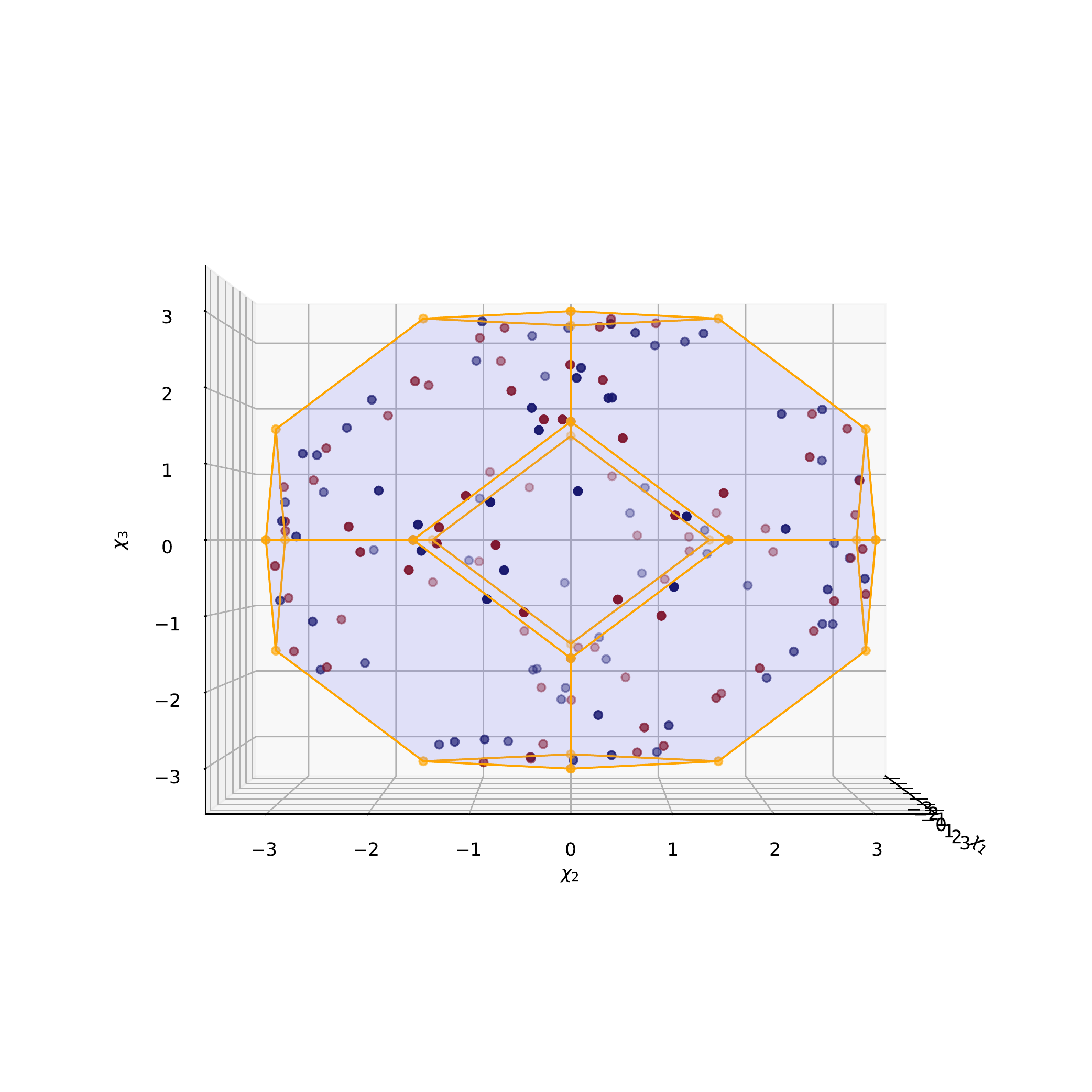}
		\caption{View angles $0,0$}
	\end{subfigure}
	\centering
	\begin{subfigure}{0.45\textwidth}
		\includegraphics[width=\textwidth]
		{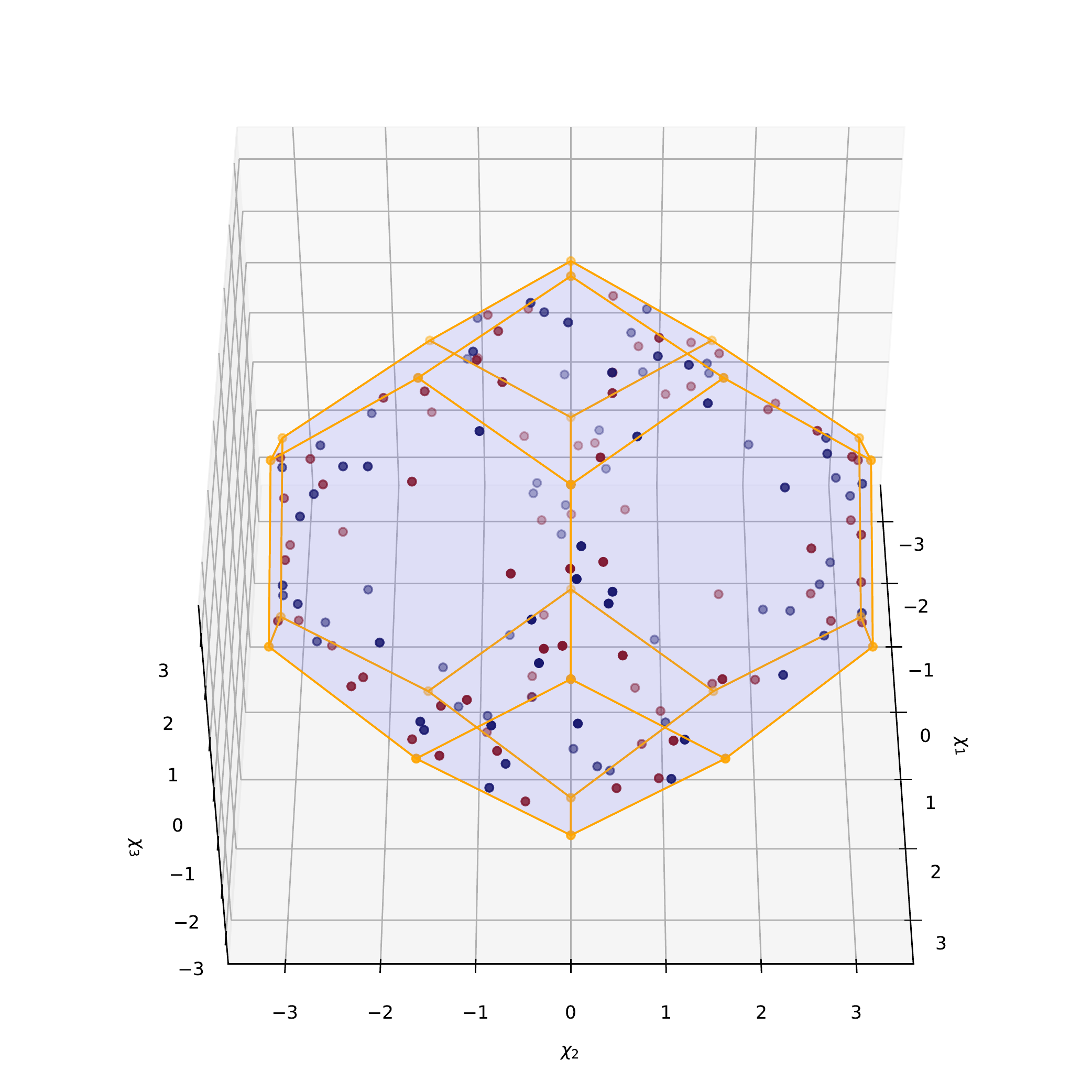}
		\caption{View angles $45,0$}
	\end{subfigure}
	\begin{subfigure}{0.45\textwidth}
		\includegraphics[width=\textwidth]
		{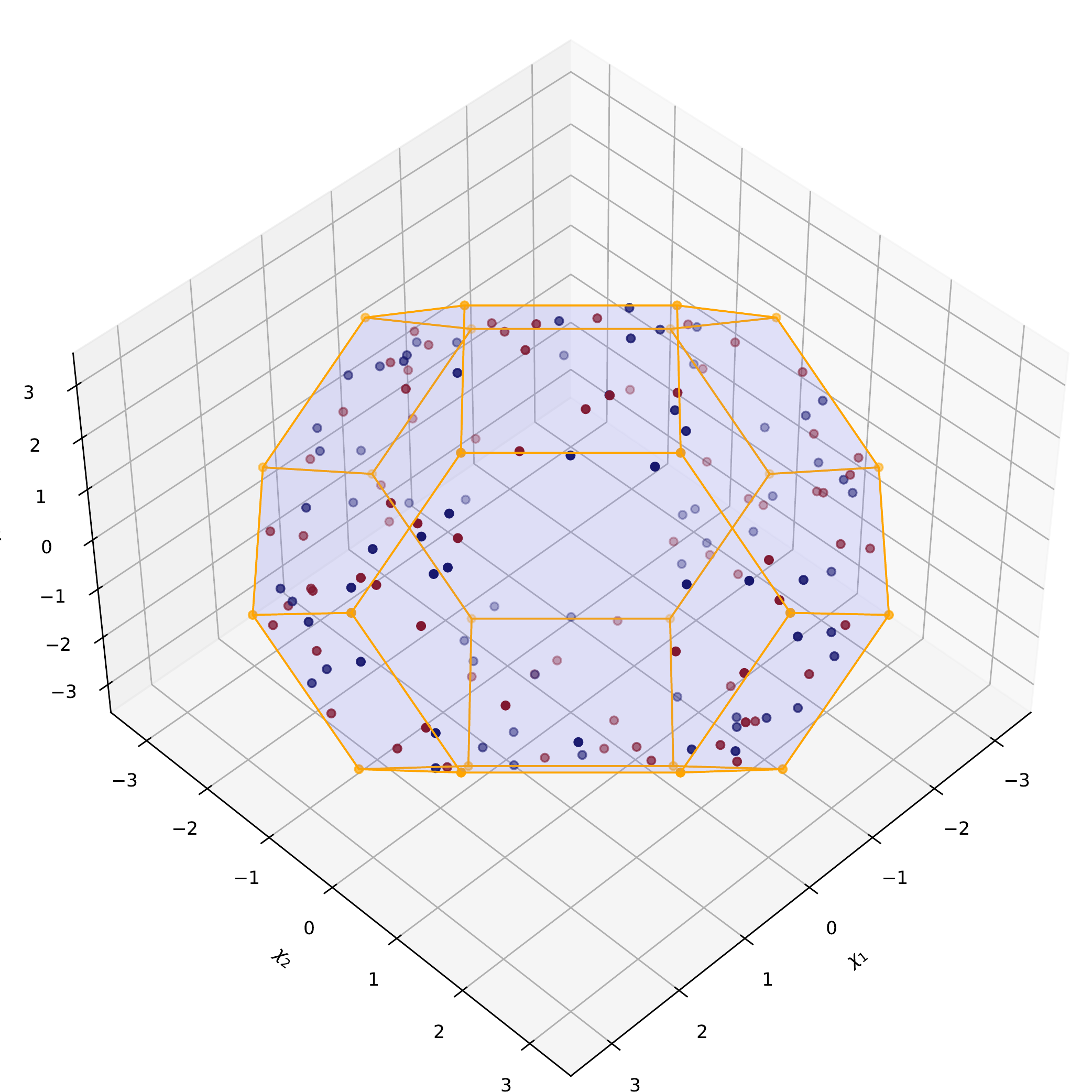}
		\caption{View angles $45,45$}
	\end{subfigure}
	\begin{subfigure}{0.45\textwidth}
	\includegraphics[width=\textwidth]
	{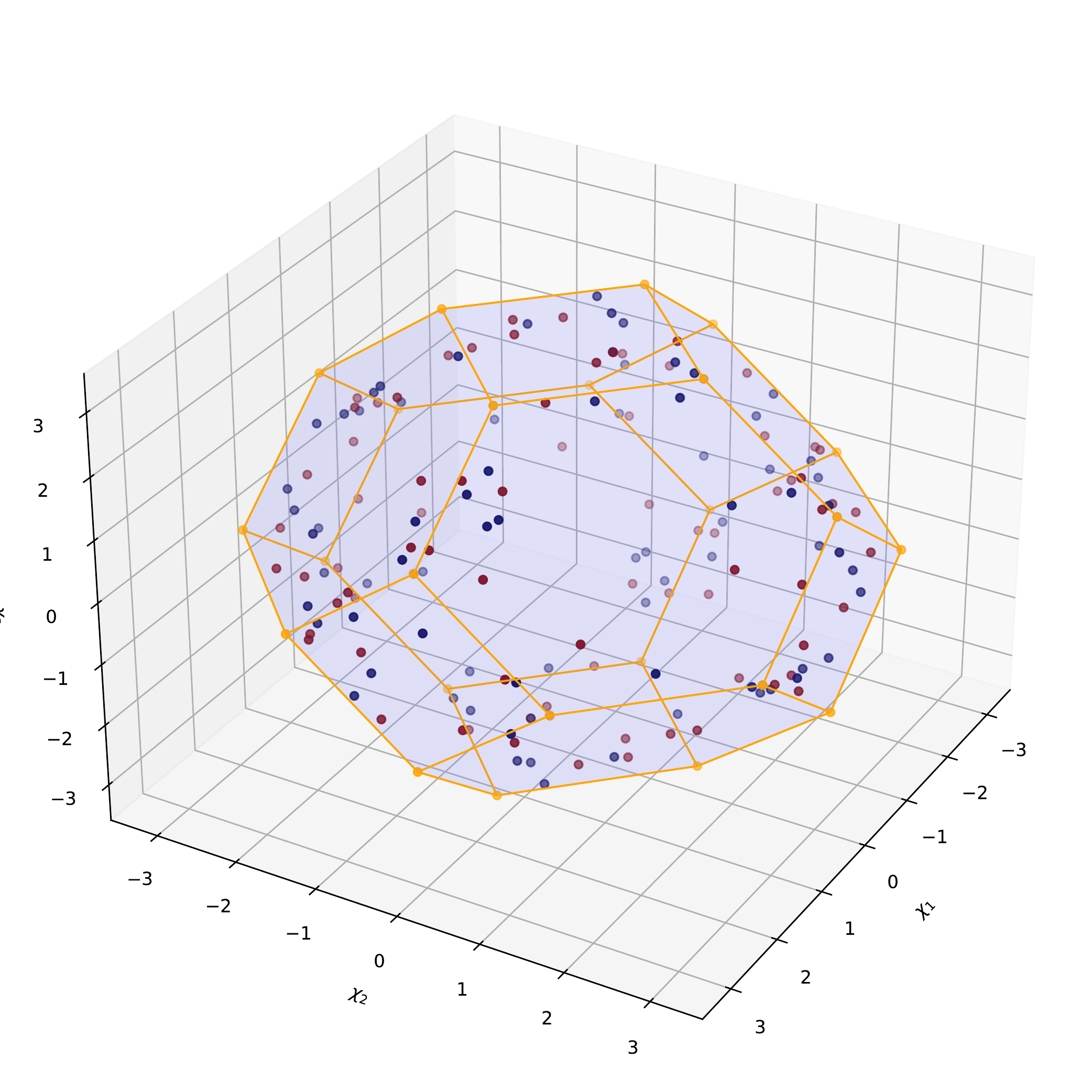}
	\caption{View angles $30,30$}
	\end{subfigure}
	\begin{subfigure}{0.45\textwidth}
		\includegraphics[width=\textwidth]
		{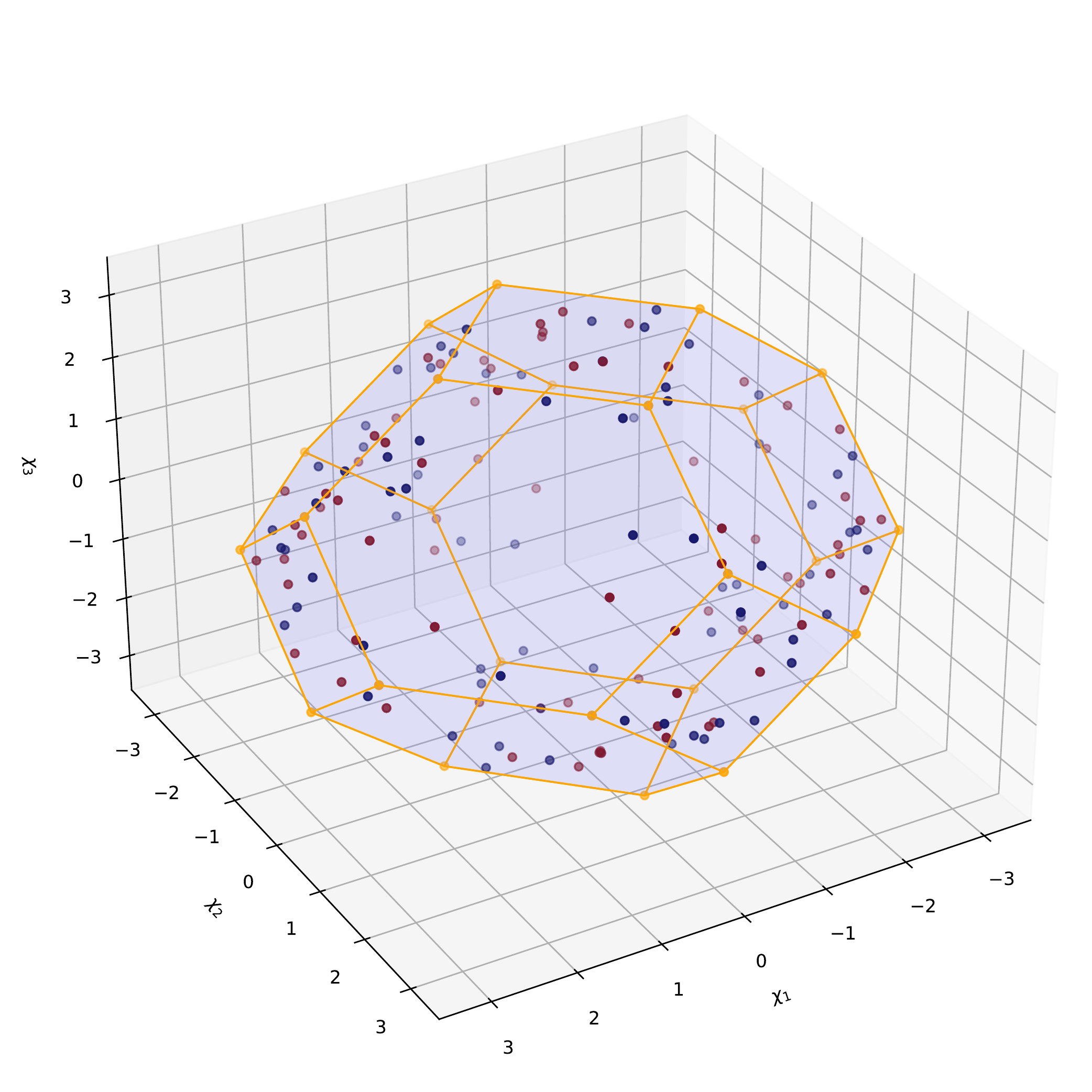}
		\caption{View angles $30,60$}
	\end{subfigure}
	\begin{subfigure}{0.45\textwidth}
		\includegraphics[width=\textwidth]
		{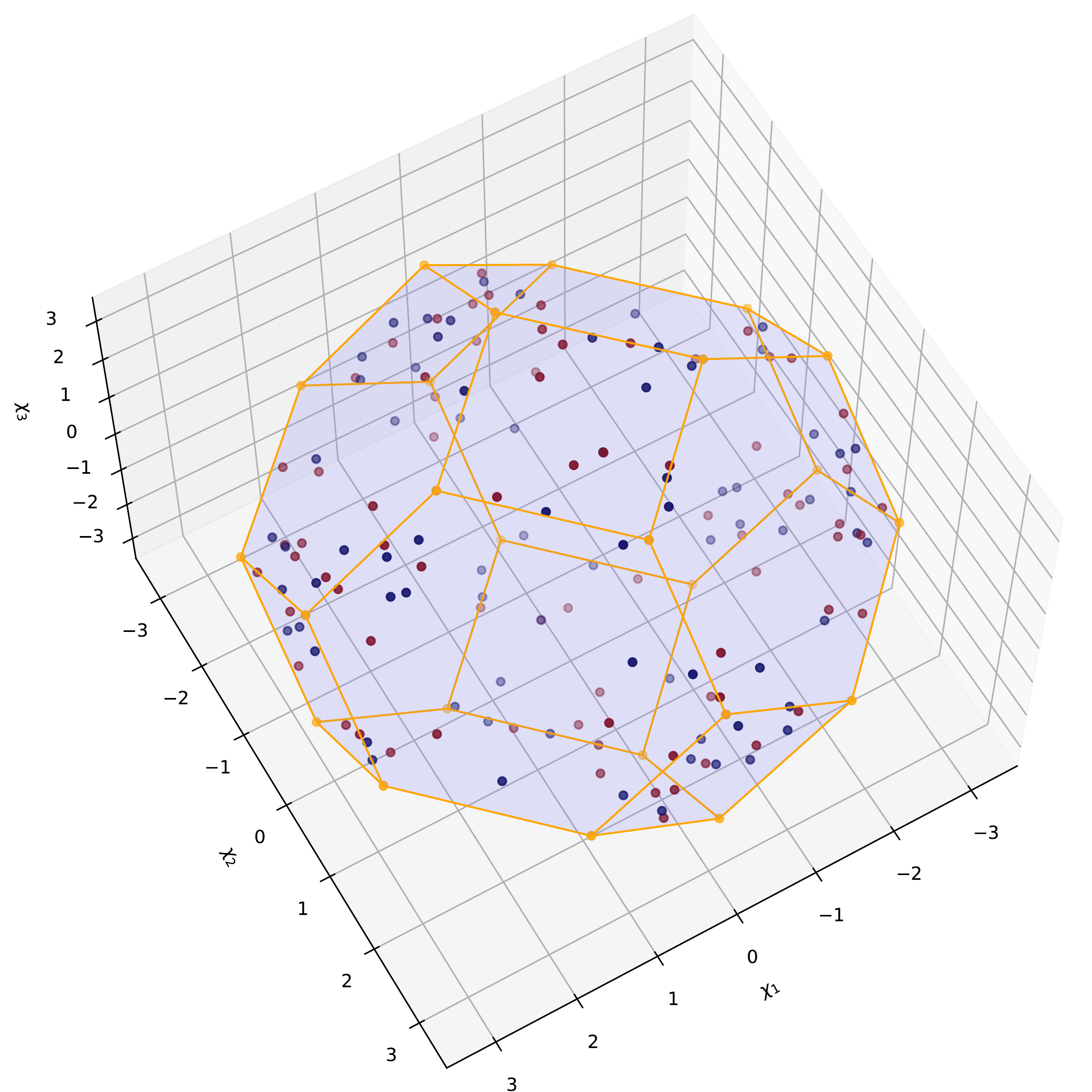}
		\caption{View angles $60,60$}
	\end{subfigure}
\caption{Positions of Weyl points for a given realization of $S$ at $N=50$ at various view angles.}
\label{fig:Supplemental-points}
\end{figure}

\end{widetext}

\end{document}